\documentclass[sn-mathphys,Numbered]{sn-jnl}

\usepackage{multirow}%
\usepackage{amsmath,amssymb,amsfonts}%
\usepackage{mathrsfs}%
\usepackage[title]{appendix}%
\usepackage{xcolor}%
\usepackage{textcomp}%
\usepackage{manyfoot}%
\usepackage{booktabs}%
\usepackage{algorithm}%
\usepackage{algorithmic}%
\usepackage{array}
\DeclareUnicodeCharacter{2212}{-}
\usepackage{listings}%
\usepackage{subcaption}  

\usepackage[english]{babel}
\addto\extrasenglish{

}    

\usepackage{setspace}

\begin{document}
\begin{titlepage}
\begin{center}
\vspace*{1cm}
\doublespacing
{ \Large Sentiment and Hashtag-aware Attentive Deep
Neural Network for Multimodal Post Popularity
Prediction}
\vspace{1.5cm}

Shubhi Bansal$^{1}$ (phd2001201007@iiti.ac.in), Mohit Kumar$^1$ (cse190001038@iiti.ac.in), Chandravardhan Singh Raghaw$^1$(phd2201101016@iiti.ac.in), \\
Nagendra Kumar$^1$ (nagendra@iiti.ac.in) \\

\hspace{10pt}

\begin{flushleft}
\small  
$^1$Department of Computer Science and Engineering,
Indian Institute of Technology Indore, Khandwa Road, Simrol, Indore 453552, Madhya
Pradesh, India.\\


\vspace{1cm}
\textbf{Corresponding Author:} \\
Shubhi Bansal \\
Department of Computer Science and Engineering,
Indian Institute of Technology Indore, Khandwa Road, Simrol, Indore 453552, Madhya
Pradesh, India.\\
Tel: +91-7988361678 \\
Email: phd2001201007@iiti.ac.in


\end{flushleft}        
\end{center}
\end{titlepage}

\title[Article Title]{Sentiment and Hashtag-aware Attentive Deep Neural Network for Multimodal Post Popularity Prediction}





\abstract{
Social media users articulate their opinions on a broad spectrum of subjects and share their experiences through posts comprising multiple modes of expression, leading to a notable surge in such multimodal content on social media platforms. Nonetheless, accurately forecasting the popularity of these posts presents a considerable challenge. Prevailing methodologies primarily center on the content itself, thereby overlooking the wealth of information encapsulated within alternative modalities such as visual demographics, sentiments conveyed through hashtags and adequately modeling the intricate relationships among hashtags, texts, and accompanying images. This oversight limits the ability to capture emotional connection and audience relevance, significantly influencing post popularity. To address these limitations, we propose a seNtiment and hAshtag-aware attentive deep neuRal netwoRk for multimodAl posT pOpularity pRediction, herein referred to as NARRATOR that extracts visual demographics from faces appearing in images and discerns sentiment from hashtag usage, providing a more comprehensive understanding of the factors influencing post popularity Moreover, we introduce a hashtag-guided attention mechanism that leverages hashtags as navigational cues, guiding the model's focus toward the most pertinent features of textual and visual modalities, thus aligning with target audience interests and broader social media context. Experimental results demonstrate that NARRATOR outperforms existing methods by a significant margin on two real-world datasets. Furthermore, ablation studies underscore the efficacy of integrating visual demographics, sentiment analysis of hashtags, and hashtag-guided attention mechanisms in enhancing the performance of post popularity prediction, thereby facilitating increased audience relevance, emotional engagement, and aesthetic appeal. 
}
\keywords{Multimodal Data Analysis, Hashtags, Sentiment, Deep Neural Network, Social Media Analysis}
\maketitle
\section{Introduction}\label{sec1}
The explosive growth of Social Network Services (SNS) such as Flickr\footnote{https://www.flickr.com/}, Facebook\footnote{https://www.facebook.com/} and Instagram\footnote{https://www.instagram.com/}, 
has revolutionized content creation and consumption \cite{shen2020examining,Anderson2021}. Users now actively generate and share vast amounts of User-Generated Content (UGC) in a wide array of formats such as texts, images, audios and videos. This has not only transformed online interactions but also resulted in a massive influx of UGC \cite{cao2020popularity} as exhibited by Flickr receiving 25 million daily uploads\footnote{https://photutorial.com/flickr-statistics/} and Instagram 1.3 billion\footnote{https://www.omnicoreagency.com/instagram-statistics/}. This abundance of content has led to information overload, making it increasingly difficult to discover relevant and engaging information. Popularity prediction emerges as a crucial solution, and aims to forecast the level of public attention UGC will receive early on. This is vital because only a small fraction of UGC truly captures widespread attention, while the vast majority remains unnoticed. By predicting popularity and identifying hot content early on, users can easily navigate information overload. The ability to forecast which content will capture public attention provides insights by identifying patterns in what resonates with their interests, revolutionize how users understand and interact with the digital world, and enable content creators to tailor their content for maximum impact. Popularity prediction models can empower platforms to make strategic decisions, optimize resource allocation, and enhance targeted advertising \cite{wu2016unfolding}. Moreover, the insights gained from popularity prediction can be leveraged to inform government policies and schemes, enhancing the effectiveness of public services. Accurate popularity forecasting has broad applications such as online marketing, and trend detection, leading to improved service effectiveness and advancements in areas such as recommender systems \cite{gonccalves2010popularity, majid2013context}, online advertising \cite{li2015click,aven2014using}, information retrieval \cite{roy2013towards,gan2016webly} and demand forecasting \cite{kim2015demand}. By analyzing patterns in what captures public attention, we gain insights into sentiment shifts and evolving user behaviors. These insights, in turn, drive advancements in sentiment analysis \cite{wang2021deep}, digital marketing \cite{saura2021using}, and user privacy protection \cite{saura2021user,ribeiro2021towards}, empowering us to make more informed decisions in the complex digital landscape.

\textbf{Unveiling Hashtag-guided Attention Mechanism}: Existing approaches to fuse multimodal data for popularity prediction 
employ attention \cite{xu2020multimodal} mechanisms to adjust feature weights dynamically. These mechanisms range from self-attention \cite{Lin2022,nguyen2019attention}, hierarchical attention \cite{liao2019popularity}, which focus on individual modalities, to mechanisms capturing interactions among modalities. For instance, context-aware attention \cite{chen2019social} captures interplay among content, time and user, dual-attention \cite{zhang2018become} considers both image-caption pairs and user environment , user guided hierarchical attention \cite{zhang2018user} learns the user's influence on texts and images, and hierarchical fusion \cite{wang2023social} operates on textual and visual modalities and attributes. However, these approaches predominantly focus on intrinsic content features, overlooking the influence of hashtags on image and text interpretation. Hashtags~\citep{Caleffi2015} i.e., words with the ``\#'' prefix, are widely used by SNS users to provide valuable contextual cues. They encode semantic relationships that reveal the creator's intended message and desired perception. Hashtags facilitate event and topic dissemination, bridging complex concepts with messages and connecting content at a granular level. Neglecting this rich source of information limits the accuracy of popularity prediction models, as hashtags significantly impact the interpretation and engagement of social media posts. Empirical evidence underscores the power of hashtags in boosting visibility since posts with hashtags garner twice the level of engagement than those without \cite{zhang2019hashtag}.
\begin{figure}[!ht]
\centering
\includegraphics[width=0.7\textwidth]{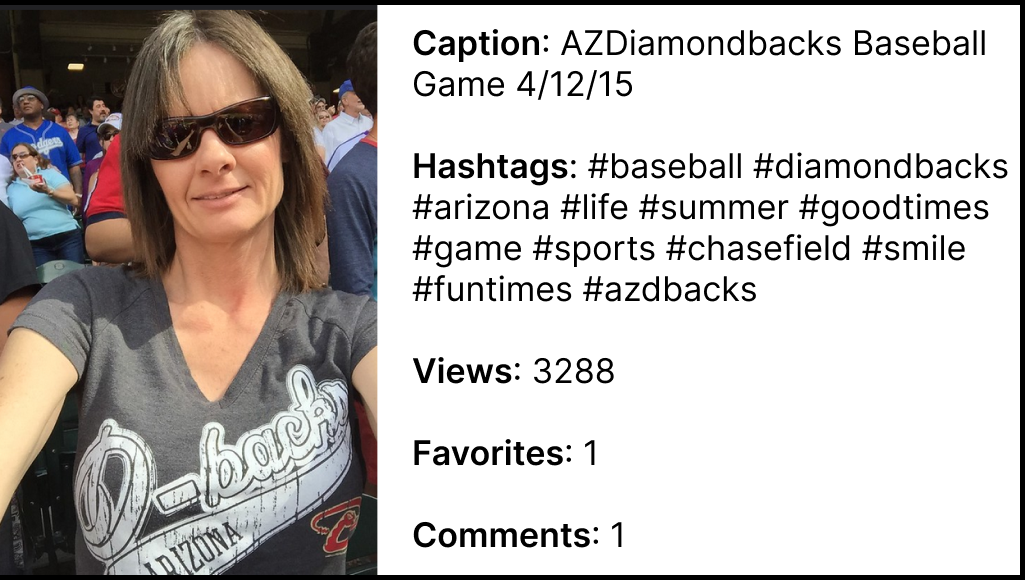}
\caption{Example Social Media Post}
\label{fig:introeg}
\end{figure} 
For instance, in the given example post (\autoref{fig:introeg}), existing attention mechanisms would analyze the selfie itself, focusing on the woman's smile, the blurred crowd in the background, or colors in her clothing. However, they might miss the contextual richness embedded in hashtags such as \#chasefield, which pinpoints the location to a baseball game at Chase Field,  \#summer which indicates the weather, and \#azdbacks and \#arizona guide the model's attention toward relevant features in the selfie: her team jersey, revealing her support for a specific team, the Arizona Diamondbacks. Therefore, it is crucial to leverage hashtags to guide model attention towards contextually relevant content features. 
In the dynamic social media landscape, where trends and user preferences evolve rapidly, such a mechanism is paramount for developing robust popularity prediction models.

\textbf{Unveiling the Power of Visual Demographics}: 
Previous studies on social media popularity prediction have largely relied on using metadata \cite{abousaleh2020multimodal} and user profiles to incorporate demographic information. However, these approaches have limitations, as metadata can be incomplete or inaccurate, and user profiles may contain outdated or intentionally misleading information. Moreover, relying solely on explicit user data can raise privacy concerns. In contrast, extracting demographic attributes directly from visual cues, particularly faces, presents an untapped opportunity. Bakshi \textit{et al.} \cite{bakhshi2014faces} found that among 1.1 million Instagram photos, those containing faces had a 38\% higher likelihood of receiving likes and a 32\% higher likelihood of receiving comments. This highlights the power of faces in capturing attention and conveying emotions, which directly impacts the post's popularity. Therefore, we aim to explore the untapped potential of visual demographic analysis in improving popularity prediction models for multimodal social media posts. By incorporating visual demographic information into popularity prediction models, we can enhance their accuracy and predictive power.

\textbf{Harnessing Sentiment from Hashtags}:
Existing methods \cite{gelli2015image,Li2019,mannepalli2023popularity} primarily focus on harnessing sentiment from the textual content of social media posts, overlooking the valuable sentiment information embedded in hashtags. Hashtags not only convey topical information but also reveal user sentiment \cite{yang2020sentiment} and audience perception, both of which can significantly influence a post's popularity. While prior studies have  leveraged structural ~\citep{liao-2022-leveraging, arazzi2023predicting} and topical information \cite{Purba2021,liao-2022-leveraging} from hashtags for predicting the popularity of multimodal content, the influence of hashtag sentiment remains underexplored.~\autoref{fig:introeg} features a user's selfie at a baseball game, annotated with hashtags such as \#baseball, \#diamondbacks, and \#arizona, conveying specific content-related information. Hashtags such as \#smile, \#funtimes, and \#goodtimes reflect the user's sentiment during the event. The sentiment expressed through hashtags can mirror the audience's collective perception of the content, providing valuable insights into ongoing conversations and prevailing trends. 
Emotional intensity within topics tends to generate more engagement, with hashtags encapsulating sentiments that might not be fully expressed in captions alone. Despite the potential of hashtag sentiment analysis for enhancing popularity prediction, it remains an underutilized resource. This research gap presents an opportunity to develop more comprehensive and accurate methods by incorporating the sentimental insights embedded in hashtags.

To address the identified research gaps, we propose NARRATOR, a Sentiment and hAshtag-aware deep neuRal netwoRk for multimodAl posT pOpularity pRediction. NARRATOR introduces a novel hashtag-guided attention mechanism allows the model to dynamically weigh the importance of different features in images and text, guided by the contextual cues provided by hashtags. This enables a more comprehensive understanding of the interplay between content and context. Furthermore, NARRATOR leverages visual cues within images to gain demographic insights, discerning fine-grained details such as age, gender, race, and emotions directly from faces. Moreover, NARRATOR explicitly incorporates sentiments extracted from hashtags, capturing the subtle emotional undertones that resonate with audiences and further refining our ability to predict post popularity. By combining these innovations- hashtag-guided attention, leveraging visual demographics, and analysing sentiment of hashtags, NARRATOR provides a deeper understanding of user engagement and emotional response, improving the performance of popularity forecasts.

Our major contributions are enlisted below.
\begin{itemize}
\item {We propose a deep neural network that leverages sentiment from hashtags, visual demographic information, and employs a hashtag-guided attention mechanism to forecast post popularity comprehensively besides content-based features and sentiment from text.}
\item{We devise a novel hashtag-guided attention mechanism that uses hashtags to guide the model's focus on content features most relevant to the intended audience and context.}
\item{Our work pioneers the use of visual demographic information for popularity prediction. We leverage visual demographics to identify engagement trends within specific audience contexts.}
\item{We derive sentiment information embedded in hashtags to decipher the emotional appeal of a post and understand how it amplifies user engagement.}
\item{Extensive experiments conducted on two real-world datasets demonstrate the superior performance of our proposed method over existing state-of-the-art methods both empirically and qualitatively.}
\end{itemize}

The rest of the paper is structured as follows. The related works are covered in~\autoref{sec:rw}. \autoref{sec:pd} presents a mathematical formulation of the problem. We elaborate on our methodology in~\autoref{sec:methodology}. The evaluations of experiments are then covered in~\autoref{sec:exp_results}.~\autoref{sec:conclusions} presents the concluding remarks of our research.
\section{Related Work}
\label{sec:rw}
In this section, we cover existing efforts in the domains of popularity prediction, hashtag representation, and attention techniques.
\subsection{Popularity Prediction in Social Media} Popularity prediction of online content has attracted significant research interest, with scholars proposing diverse methodologies. These methodologies leverage a variety of techniques, datasets, model structures, and problem formulations i.e., either classification or regression. Kumar \textit{et al.} \citep{Kumar2017} focused on predicting the popularity of news articles, particularly their ability to attract user comments, offering the opportunity for informed content modifications using various features extracted from article content.
Lin \textit{et al.}~\citep{lin2019layer} devised a framework by stacking multiple regression models across several layers, fostering synergies among diverse models to anticipate engagement of posts on a platform similar to Flickr. Purba \textit{et al.}~\citep{Purba2021} devised a novel approach for predicting Instagram post engagement rates (ER) on a global dataset. It utilizes SVR and incorporates features from hashtags, image analysis, user history, and manual image assessment. Cao \textit{et al.}~\citep {cao2020ppsp} proposed CoupledGNN, a novel graph neural network framework designed for predicting popularity with network awareness on social platforms. This framework leverages two interconnected GNNs to capture the propagation of influence. One GNN models the user activation status within the network, while the other GNN models the dissemination of information itself. Mannepalli \textit{et al.}~\citep{mannepalli2023ppm} leveraged a Long Short-Term Memory (LSTM) network to predict popularity by combining features extracted from text content, user data, time series information, and user sentiment analysis.  Furthermore, Self-Adaptive Rain Optimization (SA-RO) is employed to fine-tune the LSTM weights, enhancing the prediction accuracy. Tan \textit{et al.}~\citep{tan2022emv} leveraged transformers for time series feature extraction and CatBoost for feature selection, enabling comprehensive multi-view feature extraction and achieving superior prediction accuracy on the Social Media Prediction Dataset.

The effectiveness of these approaches depends on the selection and quality of features used to represent the content. These models may not capture several other modalities comprising the social media post, and their interrelationships might have been overlooked.
\subsection{Hashtag Representation} Prior studies have not employed hashtags used in a post as the main feature in popularity prediction tasks. According to Zappavigna~\citep{zappavigna2015searchable}, hashtags have a variety of uses on social media, with subject markers serving as the most important ones. Hashtag efficiently defines the topic of the post and help the user to easily identify if the post is related to them or not. Few researchers have also explored the representation of hashtags in the form of graphs.
Liu \textit{et al.}~\citep{liu2018hashtag2vec} proposed a network framework where hashtags serve as nodes. Each node is linked to a collection of tweets, which themselves are comprised of individual words. To capture the inherent relationships within this hierarchical, heterogeneous network, they introduce the Hashtag2Vec embedding model. This model extends its embedding capability beyond hashtags to encompass short social text elements (i.e., tweets) by simultaneously considering relationships between hashtags themselves (hashtag-hashtag), hashtags and tweets (hashtag-tweet), tweets and their constituent words (tweet-word), and finally, word-to-word relationships within tweets. Liao \textit{et al.}~\citep{liao-2022-leveraging} proposed a multimodal framework that analyzes hashtag network structure semantics, and topic modeling besides captions and images to predict popular influencer posts in Taiwan. Chakrabarti \textit{et al.}~\citep{chakrabarti2023hr} addressed the challenge of maximizing social media popularity by recommending context-relevant hashtags. The proposed framework leveraged post keywords, user popularity, and trending hashtags to recommend effective hashtags.

However, these approaches focus on the structural and textual aspects of hashtags ignoring the sentiment information embedded in them. Posts with positive or sentimental hashtags tend to perform better than those with neutral or negative hashtags. This indicates that sentiment analysis of hashtags can be employed to forecast which posts are more inclined to become viral.
\subsection{Attention Mechanism} Many researchers have stated that using the attention mechanism as a part of the popularity prediction model can significantly enhance performance. As many models use images and texts as primary features to predict the popularity score, the attention mechanism can help represent these features in a better way. Xu \textit{et al.}~\citep{xu2020multimodal} proposed a regression model to predict popularity where they used an attention layer at the top of the model and showed a significant improvement when compared to models without attention. Lin \textit{et al.} \citep{Lin2022} utilized a self-attention mechanism to fuse semantic and numeric features effectively for social media popularity prediction, outperforming other methods. 
Bansal \textit{et al.}~\citep{Bansal2023} devised a word-level parallel co-attention mechanism to derive an enriched representation of multimodal social media posts by capturing the mutual influence of text and image on each other. Wang \textit{et al.}~\citep{wang2023sm} presented a novel multimodal popularity prediction model grounded in hierarchical fusion, where extracted features encompass visual elements, textual content, along with attributes extracted from both modalities. The model innovatively integrates three integration stages- early integration, representation integration, and modality integration - culminating in a fully fused vector inputted into an XGBoost regression model for effectively estimating the virality of posts.

However, extant approaches prioritize local features, neglecting a holistic understanding of the content's multimodal nature. To overcome this limitation, we propose a novel multimodal deep learning model with an attention mechanism. This model incorporates diverse feature types known to influence post popularity, encompassing content characteristics, sentiment analysis, hashtag information, and user demographics. The proposed model uses transfer learning, deep learning, attention mechanism, and graph neural networks to learn fine-grained representations of these features and then feed these enhanced feature representations into a unified framework to estimate the post popularity.

\section{Problem Definition}
\label{sec:pd}
Suppose there are $P$ multimodal social media posts. Let $p_i$ denote the $i^{th}$ multimodal post such that $p_i= \{ p_i^t, p_i^v,p_i^d, p_i^h, p_i^m\}$. Here, $p_i^t=\{w_i^x\}_{x=1}^X$
denotes the textual modality of the post, $X$ denotes the length of the post caption and $w^x$ denotes $x^{th}$ word appearing in $p_i$. Furthermore, $p_i^v$ and $p_i^h$ denote the visual and hashtag modality of the post such that $p_i^h=\{h_i^j\}_{j=1}^H$. Here, $h_i$ is the set of hashtags associated with post $p_i$, $j$ is used to index a hashtag in set $h_i$, $H$ is the cardinality of hashtag set $(h_i)$ assigned to $(p_i)$. The symbol $p_i^d$
represents the demographic information such as age, gender, race, and emotion on the faces of people appearing in images and $p_i^{m}$ denotes metadata of $p_i$ and the user $(u_i)$ who created it which contains followers count, following count, post count, hashtag count, and caption length.
We specify our problem using the notations discussed above.

\textit{Given a multimodal social media post $(p_i)$, our aim is to train a function $f(.)$ that allows us to forecast its popularity score.}
\begin{equation}
 {\hat{y}}^i=\ f(p_i)
\end{equation}
Here, ${\hat{y}}^i$ represents the predicted popularity score for the post $p_i$. We frame the popularity prediction task as a regression problem. Our objective is to learn the enriched feature representation of $(p_i)$  and predict its popularity score. 
\section{Methodology}
\label{sec:methodology}
\begin{figure*}[htb]
	\centering
\includegraphics[width=\textwidth]{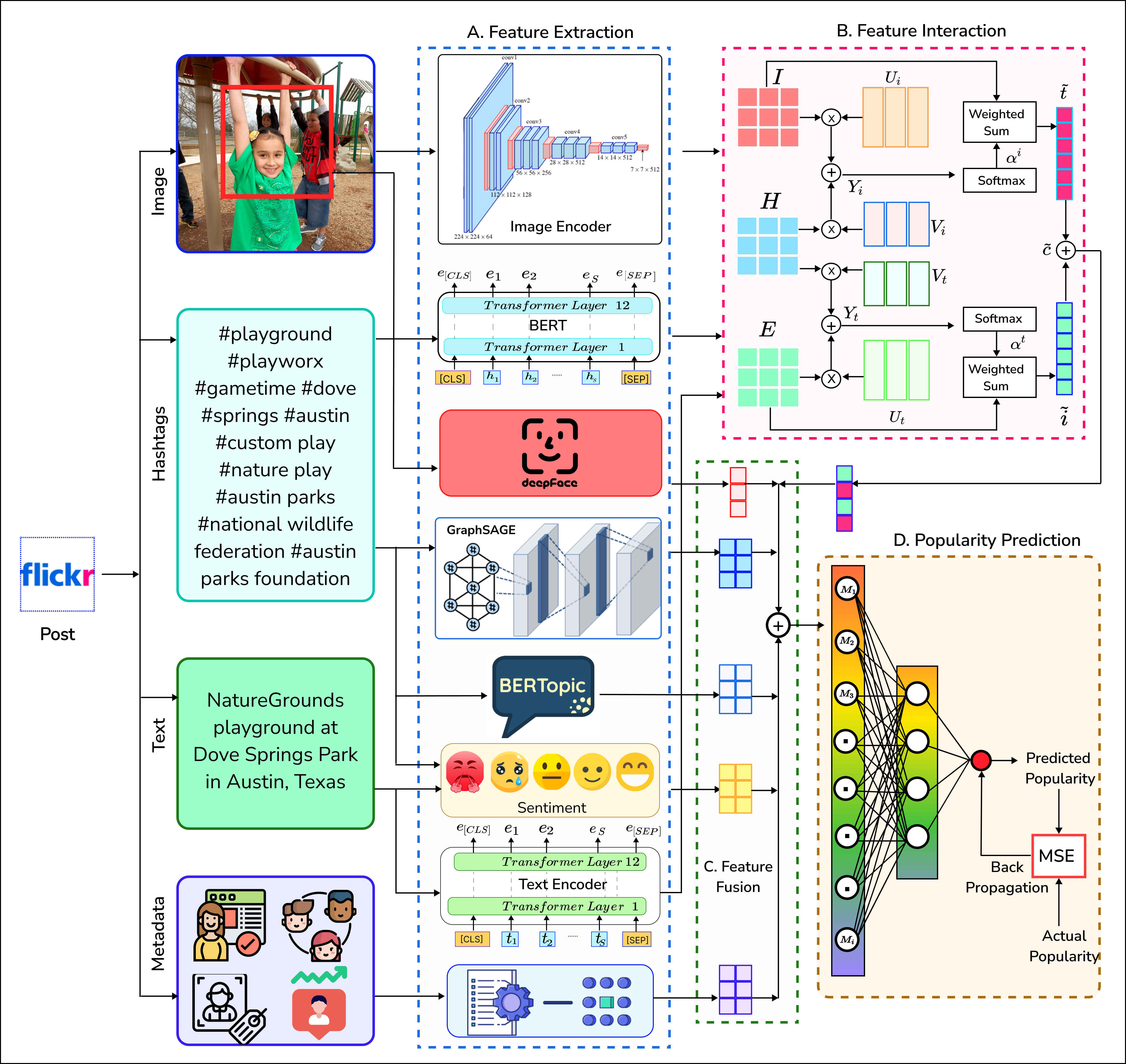}
	\caption{\centering System Architecture of NARRATOR}
	\label{fig:sa}
\end{figure*}

We introduce our novel methodological approach within this section. \autoref{fig:sa} illustrates the architecture of our proposed framework for the popularity prediction of multimodal posts. We analyze varied features that significantly influence the popularity prediction of social media content. First, we investigate the textual features of captions followed by visual features derived from images of social media posts. 
Additionally, we examine the demographic information from the faces of people appearing in images accompanying social media posts. Then, several social features based on user and post metadata are explored. We leverage content-based and sentiment-based information from hashtags and post captions to effectively capture the rich information embedded in these posts. We derive topical and structural information from hashtags annotated to these posts. We also learn the mutual influence of hashtags on visual and textual modalities by devising a novel hashtag-guided attention mechanism. These features are then passed to several dense layers to predict the popularity score. Our proposed framework entails a three-step process for accurately predicting the popularity score of social media posts: (1) feature extraction, (2) feature interaction, and (3) feature fusion for popularity prediction. These steps are discussed in detail below.
\subsection{Feature Extraction}
In this part of the paper, we discuss the feature retrieval procedure for multimodal posts.
\subsubsection{Textual Feature Extraction}
Social media posts inherently rely on user-provided captions for context. To extract a textual feature representation from these captions, we leverage a transformer-based model known as Bidirectional Encoder Representations from Transformers (BERT)~\cite{devlin-etal-2019-bert}. Limited by its context-agnostic approach, word2vec~\cite{mikolov2013distributed} cannot effectively handle homonyms. BERT, on the other hand, prioritizes the words surrounding a target word during the embedding creation process. This enables BERT to capture the nuances of language and generate more semantically rich representations.

For the textual modality of social media post ($p_i^t$) which comprises a word sequence denoted by $p_i^t=\{w_i^x\}_{x=1}^W$, we add two tokens: class (CLS) and separator (SEP) to indicate the beginning and end of the input text respectively. Here, $W$ is the number of words appearing in the post caption. We generate the corresponding set of tokens $T$ using the BERT tokenizer as given in~\autoref{eq:eqtfe1}.
\begin{equation}
T = BERT\_T okenizer(p_i^t) 
\label{eq:eqtfe1}
\end{equation}
We process the text sequence, denoted by $T$, through the BERT model as defined in~\autoref{eq:b}. This process yields a 768-dimensional vector representation for each token within the sequence.
\begin{equation}
    B = BERT(T)
    \label{eq:b}
\end{equation}
Here, $B = \{e^x\}^M_{x=1}$ is a matrix that encodes the textual features extracted from the post caption using BERT. This matrix comprises $M$ rows, where $M$ represents the fixed length of the token sequence. Each row $e^x$ contains the 768-dimensional BERT embedding for a corresponding token within the sequence (where $x$ denotes the token index ranging from 1 to M). For textual descriptions less than $M$, we apply padding, otherwise, we perform truncation to make all textual descriptions of
uniform size. Further, we use Long Short Term Memory (LSTM) to model the sequential relationship among words. The LSTM unit outputs a hidden state $t_i^x$ for the current word $w_i^x$ by taking the embedding of the current word derived from BERT i.e., $e_i^x$ and the hidden state of the preceding time step $t_i^{x-1}$ as inputs as shown in \autoref{eq:tf2}.
\begin{equation}
t_i^x = LSTM(e_i^x, t_i^{x-1})
\label{eq:tf2}
\end{equation}
Here $t_i^x \in \mathbb{R}^D$, $x = 1, 2,\hdots, M$, and $D$=768. For the sake of conciseness, we skip the specific LSTM formulae. We stack hidden state feature vectors for each word derived from LSTM to generate textual feature matrix $E$ as given in~\autoref{eq:eqm2}.
\begin{equation}
E = \{t_i^x\}_{x=1}^M
\label{eq:eqm2}
\end{equation}
Here, $E \in \mathbb{R}^{M \times D}$ is the textual feature matrix, and $M = 15$ denotes the maximum length of the associated text for the post. The dimension of each $t_i^x$ is $\mathbb{R}^D$ where $D=768$ denotes the embedding dimension.
\subsubsection{Visual Feature Extraction}
The image of the social media post plays a pivotal role in predicting the post's popularity. Deep learning approaches for extracting visual information have progressed remarkably in recent years. To extract visual features of the post, we employ VGG19~\citep{simonyan2014very} model. The VGG19 model classifies 1.2 million images from the ImageNet~\citep{deng2009imagenet} database into 1000 categories during its training process. We extracted visual features using the output of the final pooling layer of VGG19. We create several feature vectors for a picture by retaining the regional feature vectors. The feature matrix for an image $(V)$ can be expressed as exhibited in~\autoref{eq:vf2}.
\begin{equation}
V = \{v_i^k\}_{k=1}^K
\label{eq:vf2}
\end{equation}
 Here $v_i^k \in \mathbb{R}^N$, $k = 1, 2,\hdots, K$ with $N = 512$ which denotes the size of regional feature vector. We retain $K = 7 \times7= 49$ regional feature vectors for each image since the final pooling layer of VGG-19 is a $7\times7\times 512$ tensor for $7\times7$ regions, each of which is represented by a 512-dimensional vector. Following the feature extraction stage, a fully connected (FC) layer is employed to project each regional feature vector into a new vector space. This transformation ensures that the dimensionality of the resulting image feature vectors aligns with the dimensionality of the text feature vectors, facilitating their subsequent concatenation and joint processing within the model architecture. The mathematical formulation for this transformation is presented in~\autoref{eq:vf1}.
\begin{equation}
    I =\{v_i^k\}_{k=1}^K 
    \label{eq:vf1}
\end{equation}
Here, $I$ is the visual feature matrix where $I \in \mathbb{R}^{K \times D}$ and $v_i^k \in \mathbb{R}^D$ where $D=768$ is the embedding dimension for each regional feature vector.
\subsubsection{Demographic Feature Extraction}We used DeepFace~\citep{serengil2021hyperextended}, a compact framework for identifying faces and analyzing characteristics namely age, gender, race, and emotion from faces of people appearing in images associated with the uploaded post. The VGG-Face model was used to investigate DeepFace. In DeepFace, 101 nodes are present for predicting the age between 0 to 100 years of the person present in the image. The race model predicts six different races namely Black, White, Asian, Middle Eastern, Indian, and Latino. The emotion on the users’ faces is computed as one of the seven categories i.e., fear, sadness, happiness, anger, disgust, surprise, and neutral. The gender of the user is defined as male or female. Race, emotion, and gender were present as integral values. At last, we derive the demographic feature vector by concatenating gender, age, emotion, and race as given in~\autoref{eq:eqm4}.
\begin{equation}
f_i^{d}=\{g, a, e, r\}
\label{eq:eqm4}
\end{equation}
Here, $f_i^{d}$ denotes the derived demographic feature vector that has a dimension of 116 and $g, a,e,r$ represents gender, age, emotion, and race. 
\begin{figure}[H]
\centering
\includegraphics[width=\textwidth]{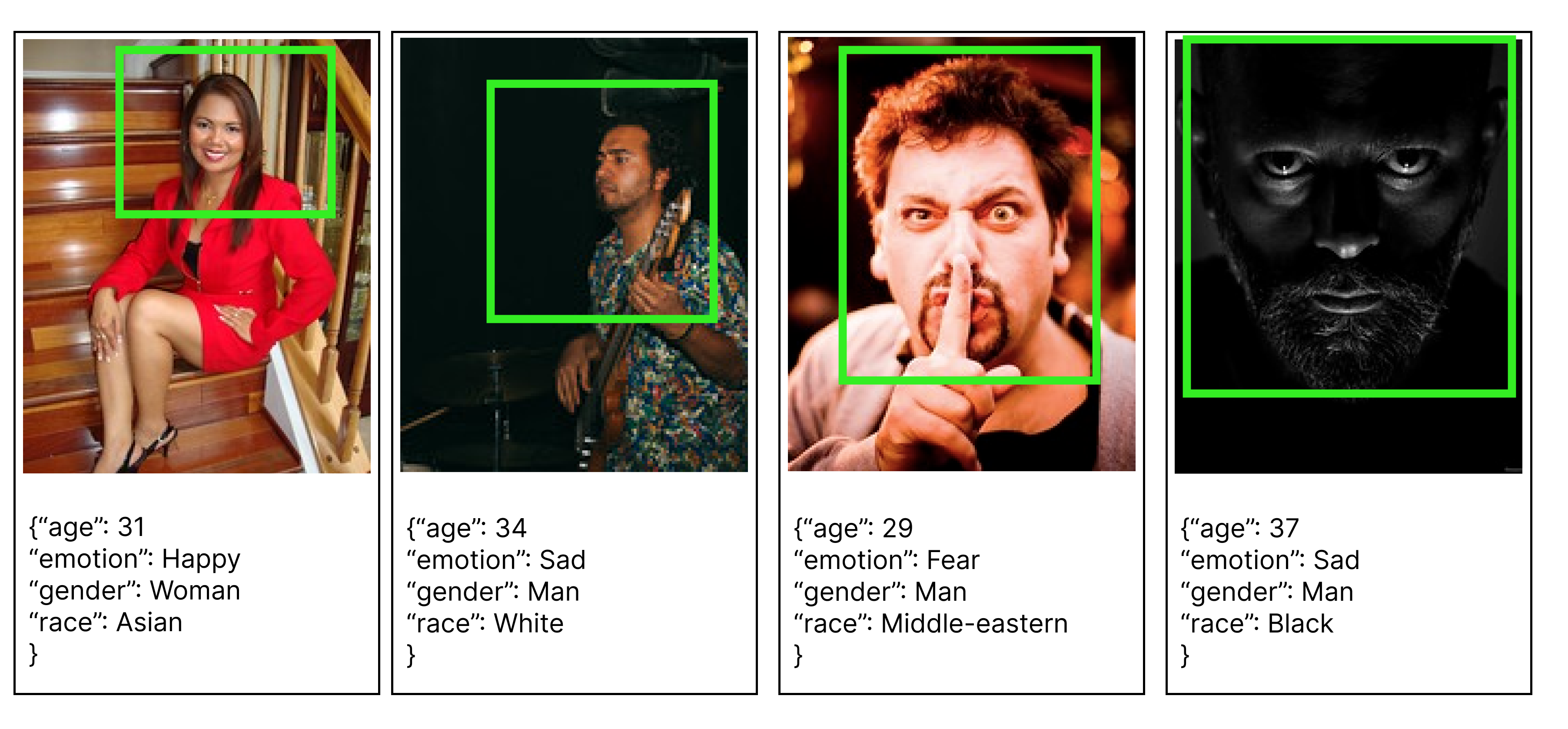}
\caption{Posts depicting Demographic Features}
\label{fig:demographic}
\end{figure}
~\autoref{fig:demographic} shows shows four Flickr posts where we used a DeepFace model to analyze facial features and infer emotions and demographics. The first Flickr post shows an Asian woman with long hair who appears to be in her early thirties, smiling and happy. The second post shows an early middle-aged white man playing guitar and is sad. The third post shows a man with eyes frowning and trying to silence; hence, the
inferred emotion from facial expression is fear. The fourth post is a man who is in his late thirties, inferred ethnicity is black, and emotion is sad because of his furrowed brows, downturned mouth, and dimly lit surroundings.
\subsubsection{Hashtag Feature Extraction}
In social media, hashtags serve as useful subject labels and search tools. Rather than attractive titles or pictures, trendy hashtags are the reason behind some social media posts getting a lot of attention. Hashtags encapsulate important information that should be incorporated in social media post representation. However, conveying textual or topical information is just as crucial as providing the hashtag network's structural components.
Hence, we define the hashtag feature as a combination of topic embeddings and node embeddings that represent the content of the hashtag and the graphical structure between hashtags respectively.
For the extraction of topic embeddings $(X_h)$, we used the BERTopic~\citep{grootendorst2022bertopic}. BERTopic is a topic modeling technique that builds topic representations using the transformers framework and c-TFIDF. BERTopic first uses sentence transformers to produce a number of document embeddings. Next, it uses HDBSCAN~\citep{mcinnes2017hdbscan} for document clustering and UMAP~\citep{mcinnes2018umap} for embedding dimension reduction. In order to determine the relevance of each word inside each subject, we compute the class-based TFIDF for every cluster (topic).
The average of all document embeddings inside a given subject is used to determine the topic embedding for a particular topic.

For the structure embedding, we constructed a graph-like network where we used hashtags as nodes of the network, and edges between two hashtags are created based on co-occurrence. We assign weights to the edge as the number of times the two hashtags appear in a single post. Further, we calculated structure embeddings $\bar{V}_h$ for each node using GraphSAGE~\citep{hamilton2017inductive}. GraphSAGE is used for inductive representation learning on huge graphs. When creating low-dimensional vector representations of nodes, GraphSAGE is particularly helpful for graphs that include a wealth of node attribute data. The structural embedding of a post denoted by $\bar{V}_h$ is determined only if there are at least two hashtags present in the post and taking the average of node embeddings of hashtags appearing in the post. On the other hand, a zero-vector is allocated as the structural embedding if a post is devoid of any hashtags. We then concatenate the topic and structure embedding to derive the overall hashtag representation for the post as shown in~\autoref{eq:hf1}.
\begin{equation}
f_i^{h}= \{X_h  \oplus \bar{V}_h\}
\label{eq:hf1}
\end{equation}
Here, $f_i^{h}$ is the resultant hashtag feature vector post $p_i$, $X_h$ is the topic embedding and  $\bar{V}_h$ is the average hashtag node embedding. 
\subsubsection{Sentiment Feature Extraction}
We embed each post caption into a 5-dimensional vector using Stanford's CoreNLP Sentiment Analysis tool\footnote{https://stanfordnlp.github.io/CoreNLP/}. The scale for sentiment values ranges from zero to four which represents the likelihood that the sentence is extremely negative, negative, neutral, positive, or very positive. This tool was created by the Stanford NLP group as a module of the Stanford CoreNLP toolset~\citep{manning2014stanford}. Java is used to power Stanford's CoreNLP. Unlike Vader~\citep{hutto2014vader} and TextBLob which look at the sentiment of individual words, Stanford CoreNLP output the sentiment values based on the entire sentence structure resulting in improved performance.
We derive the sentiment features from the post caption as shown in~\autoref{eq:s1}.
\begin{equation}
s^t=Stanford\;CoreNLP(p_i^t)
\label{eq:s1}
\end{equation}
Here, $s^t$ denotes the sentiment feature vector derived from the textual modality of the post ($p_i^t$). We additionally treat hashtags as sentences and construct a 5-dimensional vector from the hashtag modality ($p_i^h$) of the multimodal social media post using Stanford CoreNLP.
\begin{equation}
s^h=Stanford\;CoreNLP(p_i^h)
\label{eq:s2}
\end{equation}
Here, $s^h$ denotes the hashtag-based sentiment feature vector having a dimension of 5. The overall sentiment feature vector for the post ($p_i$) is derived by concatenating the text-based and hashtag-based sentiment feature vectors as illustrated below.
\begin{equation}
f_i^{st}=concat(s^t,s^h)
\label{eq:s3}
\end{equation}
Here, $f_i^{st}$ is the resultant sentiment feature vector for the post ($p_i$) having a dimension of 10, $s^t$ and $s_h$ denote the text-based and hashtag-based sentiment feature vectors, respectively.
\subsubsection{Social Feature Extraction}
The multimodal post's popularity is influenced by both its content and the user who posted it in terms of social media presence~\citep{aloufi2017prediction}. We have categorized social features into two categories i.e., user metadata and post metadata. We discuss these two below.
\begin{enumerate}
\item User Metadata: The number of prior posts a user has made and their activity on the platform are both strongly connected with the popularity of their most recent post. Therefore, we have taken some user-centric features which are as follows:
\begin{enumerate}
    \item User Id: A unique integer defining the user on the platform uniquely.
    \item Average Views: It is obtained by computing the sum of all views over all the posts posted by the user in the past divided by the number of his previously uploaded posts.
    \item Group Count: Total number of groups the user has joined on that platform.
    \item Average Member Count: Average number of members in the group that the user joined.
 \end{enumerate}
\item Post Metadata: The textual information associated with a post significantly influences the post's popularity. A post with a large title may not get huge popularity or a post with a large number of hashtags will appear more frequently so that it may gain more popularity.
The post metadata consists of:
\begin{enumerate}
\item Tag Count: The number of hashtags a person used in their post.
\item Title Length: Word count of the caption of the post.
\item Description Length: Length of the description of the post
\item Tagged People: It is defined by a binary number 0 if people are not tagged in the post else 1.
\item Comment Count: The number of comments received by a post from other users.
\end{enumerate}
\item{Time}: Beyond user and content characteristics, predicting post popularity necessitates incorporating temporal features. Research suggests a diurnal cycle in social media activity, with weekends experiencing increased user engagement. Consequently, posts uploaded during these high-activity periods are tend to garner more views and interactions. To account for this temporal influence, we leverage the following time-based features:
\begin{enumerate}
\item{Post Day:} This categorical feature denotes the day of the week on which a post is uploaded. We employ one-hot encoding to represent the post-day as a 7-dimensional vector.
\item{Post Month:} This categorical feature indicates the month in which a post is uploaded. Similar to post-day, one-hot encoding is used to represent the post-month as a 12-dimensional vector.
\item{Post Time:} This categorical feature captures the time of day during which a post is uploaded. We divide the day into four distinct time segments (morning, afternoon, evening, and night), each encompassing six hours. One-hot encoding is then applied to represent the post time as a 4-dimensional vector.
\item{Post Duration:} This numerical feature represents the number of days an image remains posted on the platform (e.g., Flickr).
\end{enumerate}
\end{enumerate}
By incorporating these temporal features, our model can learn how time-related trends impact post popularity, potentially leading to more accurate predictions.
The social feature vector ($f_i^{s}$) is obtained using the user ID, average views, group count, average member count, tag count, title length, description length, tagged people, comment count, and temporal data.
\subsection{Feature Interaction}
The feature interaction module sheds light on how hashtags interact with textual and visual modalities by devising a novel hashtag-guided attention mechanism. At its core, this mechanism utilizes hashtags as guiding signals to focus the attention of the predictive model on relevant features within the content. By considering hashtags associated with the content, the model can better understand the context in which the content is shared, leading to more accurate predictions.
It allows the model to adapt its focus dynamically, making it suitable for a wide range of content types and social media platforms. Algorithm~\autoref{alg:alg1} shows how our devised hashtag-guided attention mechanism leverages hashtags to guide attention toward relevant text and image features. The use of hashtag embeddings and attention weights provides insights into the factors influencing content popularity, making the model more interpretable for users and content creators. By doing so, it aims to improve the prediction of social media post popularity.
\begin{algorithm}
\caption{Hashtag-guided Attention}
\label{alg:alg1}
\begin{tabular}{ll}
    \textit{Input:} & $E$: Text feature matrix\\
    & $V$: Image Feature Matrix\\
    & $H$: Hashtag Feature Matrix\\
  \textit{Output:} & ${\tilde{c}}$:  Updated Content Feature Vector\\
  \vspace{0.05in}
  \textbf{function } &  Hashtag-guided Attention($T, V, H$)
    \end{tabular}
\begin{algorithmic}[1]
\FORALL{$t = 1$ \TO $M$}
    \STATE $Y_t[t] = \tanh(E[t] \times U_t + H \times V[t])$
\ENDFOR
\FORALL{$i = 1$ \TO $K$}
    \STATE $Y_i[i] = \tanh(V[i] \times U_i + H \times V[i])$
\ENDFOR
\FORALL{$t = 1$ \TO $M$}
    \STATE $\alpha^t[t] = \text{softmax}(Y_t[t] \times W_t)$
\ENDFOR
\FORALL{$i = 1$ \TO $K$}
    \STATE $\alpha^i[i] = \text{softmax}(Y_i[i] \times W_i)$
\ENDFOR
\FORALL{$j = 1$ \TO $D$}
    \STATE $\tilde{t}[j] = \sum_t (E[t][j] \times \alpha^t[t])$
\ENDFOR
\FORALL{$j = 1$ \TO $D$}
    \STATE $\tilde{i}[j] = \sum_i (V[i][j] \times \alpha^i[i])$
\ENDFOR
\STATE $\tilde{c} = \tilde{i} + \tilde{t}$
\RETURN $\tilde{c}$
\end{algorithmic}
\label{alg1}
\end{algorithm}
Lines 1-6 show how to compute the intermediate representation of text and image based on hashtags. We apply transformations on text and image feature matrices using learnable parameters to capture their interactions with hashtags. Here, $Y_t \in \mathbb{R}^{D \times A} $ and
$Y_i \in \mathbb{R}^{D \times A}$ are the intermediate representation of the text and image feature matrix based on hashtags, respectively, $A$ is the number of attention units set to 768, $U\textsubscript{t} \in \mathbb{R}^{M \times A}$, $V\textsubscript{t} \in \mathbb{R}^{L \times A}$, $U\textsubscript{i} \in \mathbb{R}^{K \times A}$, $V\textsubscript{i} \in \mathbb{R}^{L \times A}$ are learnable parameters. 
The hashtags associated with the post are embedded into a continuous vector space representation by using BERT. 
\begin{equation}
H=BERT(p_i^h)
\end{equation}
Here, $H \in \mathbb{R}^{L\times D}$ is the resultant hashtag feature matrix, $L$=60 based on the maximum number of hashtags associated with a post. To have a feature matrix of uniform dimensions across different posts in the data, we padded zeros for posts having a tag count of less than 60. We employ BERT because one hashtag can have different meanings in different posts. For example, \#rock can be used to refer to stones and in other posts, the same hashtag can refer to music rock band. Therefore, it is important to capture the context in which a particular hashtag is being used. BERT-based embeddings of hashtags capture their semantic relationships, allowing the model to understand their contextual meanings. By introducing learnable parameters associated with the transformation of text/image features, the model can adaptively learn how to combine these features with hashtag features to derive meaningful representations. This allows the model to adapt to the specific characteristics of the content and the nuances of hashtag usage patterns. The resulting intermediate representations encapsulate not only the inherent characteristics of the text/image features but also their contextual relevance concerning the associated hashtags. This semantic enrichment enhances the model's ability to understand the underlying themes, topics, and sentiments expressed in the content, thereby improving the quality of feature representations. 
These intermediate representations $Y_t$ and $Y_i$ signify how text and image features are influenced by associated hashtags. We apply the hyperbolic tangent (tanh) activation function to a combination of text feature matrix $(E)$ and hashtag feature matrix $(H)$ represented by $Y_t$.
Similarly, for image features, we apply tanh to a combination of image feature matrix $(I)$ and hashtag feature matrix $(H)$ which is denoted by $Y_i$. The intuition here is to capture the interaction between text and hashtags $(Y_t)$, and image and hashtags $(Y_i)$. Lines 7-12 compute the attention weights for text and image features. Here, $\alpha^t$ and $\alpha^i$ denote attention weights for text modality and image modality, respectively. 
Here, $W\textsubscript{t} \in \mathbb{R}^{A \times D}$, $W\textsubscript{i} \in \mathbb{R}^{A \times D}$ are learnable parameters. 
These weights represent the importance of different features based on the associated hashtags. The idea is that certain hashtags may be more relevant to either text or image content, affecting their contribution to the overall content vector. Then, we compute the attended modality representations. Here, 
$\tilde{t} \in \mathbb{R}^D$ and 
$\tilde{i} \in \mathbb{R}^D$ denotes the attended text feature vector and attended image feature vector, respectively.
In Lines 13-15, we take the weighted sum of text feature matrix $(E)$ with the attention weights $\alpha^t$ to get an attended text representation $\tilde{t}$.
This step emphasizes the text features that align with relevant hashtags. 
Similarly, in Lines 16-18, we multiply the image feature matrix $I$ with the attention weights $\alpha^i$ to get an attended image feature vector representation ($\tilde{i}$). Here, the focus is on image features associated with specific hashtags.
Line 20 computes the hashtag-guided content feature vector by taking the sum of the attended text feature vector $\tilde{t}$ and the attended image feature vector $\tilde{i}$. Here, $\tilde{c} \in \mathbb{R}^D$ represents the comprehensive feature vector derived from hashtag guidance. This updated feature vector represents a comprehensive view of the content, incorporating information from both text and image modalities, with attention focused on the relevant features guided by hashtags. By leveraging hashtags as guiding signals, the algorithm enhances the model's ability to capture contextually relevant features, ultimately improving the performance of the model in predicting popularity.
\subsection{Feature Fusion}
In this section, we delve into the details of feature fusion, a critical step in constructing a unified representation of multimodal social media posts and ultimately predicting their popularity within the proposed framework followed by the theoretical background for feature fusion.
\begin{figure*}[ht]
	\centering
\includegraphics[width=\textwidth]{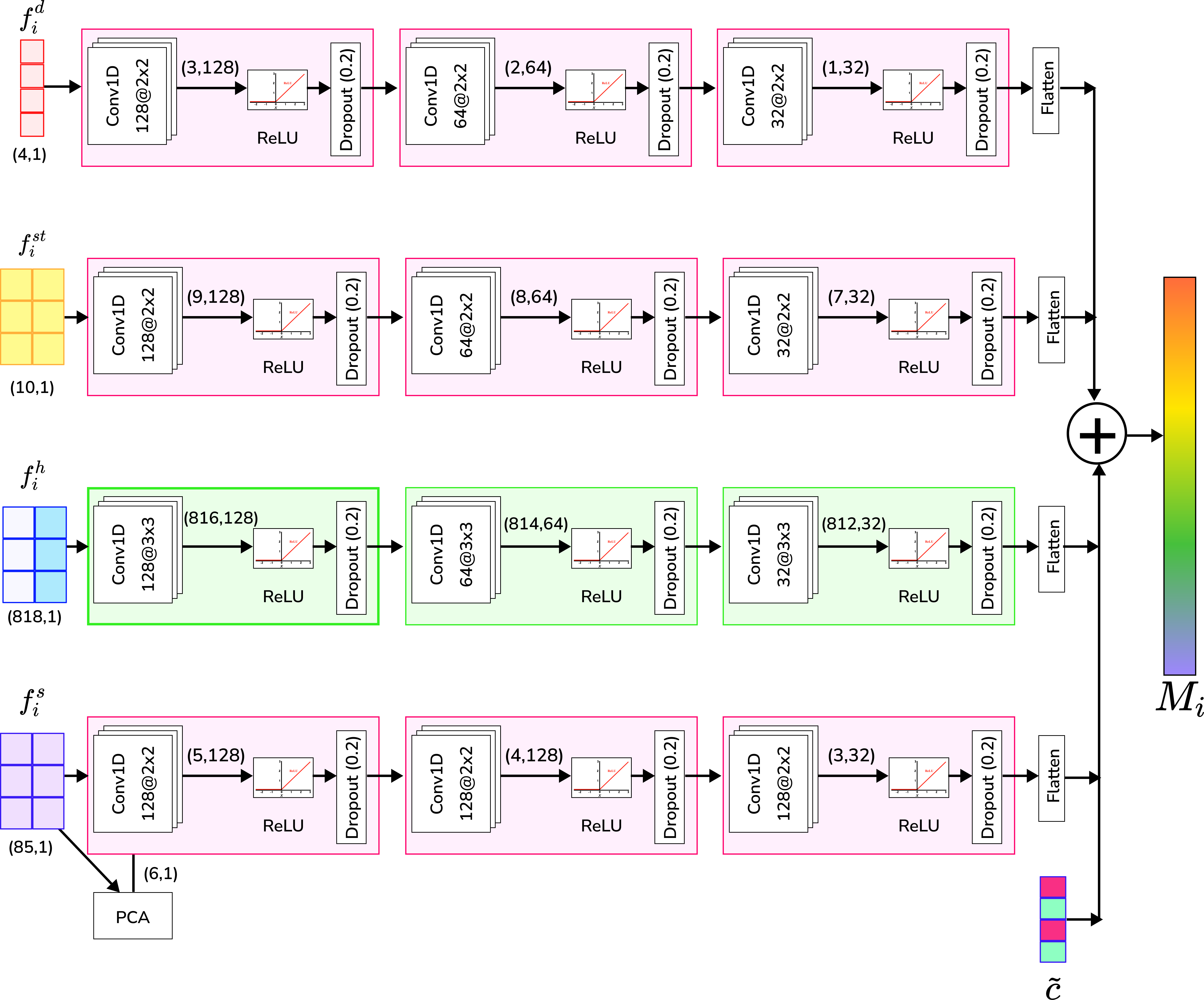}
	\caption{\centering Feature Fusion}
	\label{fig:fusion}
\end{figure*}
\autoref{fig:fusion} depicts the underlying details of the feature fusion component which is structured in a grid-like manner, with layers arranged horizontally. Each layer corresponds to a stage in the data processing conducted by the CNN. During the feature extraction stage, we extract demographic, sentiment, hashtag, and social context features from each post denoted by $f_i^d, f_i^{st}, f_i^h, f_i^s$, respectively. Conv1D refers to a one-dimensional convolutional layer, a common building block in CNNs for processing sequential data. Each layer has a box with parameters such as the filter size, number of filters, and activation function (ReLU). These parameters define how the layer performs its operations on the data. Dropout layers randomly set a fraction of activations to zero during training. This helps prevent the network from overfitting to the training data.
After the convolutional layers, there are ``Flatten” layers. These layers flatten the data from a multi-dimensional representation into a one-dimensional vector suitable for feeding into a fully connected layer.

Following the extraction of social features, we concatenate them into a single feature vector with a dimensionality of 85. To address potential issues of high dimensionality and redundancy within this feature space, we employ Principal Component Analysis (PCA)~\citep{jolliffe2016principal}. PCA is a widely recognized method for reducing dimensionality, wherein the data is projected into a lower-dimensional space to maximize the retained variance.

PCA is a well-established dimensionality reduction technique that transforms the data into a lower-dimensional space while maximizing the captured variance. In this context, we utilize PCA to reduce the dimensionality of the social feature vector from 85 to 6, thereby selecting the most salient features that contribute most significantly to the prediction task. For demographic features, we obtain a feature vector of dimension four. As stated above the final hashtag vector has a dimension of 818. The combined sentiment feature vector derived from hashtags and captions has a dimension of 10.
All feature vectors thus obtained are passed to an individual network consisting of three CNN layers. The output of all CNNs along with the hashtag-guided content feature vector is fed into a fusion network.
The fusion network is composed of a merged layer and a series of several Fully Connected (FC) layers termed Cascade Feed-Forward Network (CFFN). The output of all CNN networks along with the hashtag-guided content feature vector are concatenated in the merged layer. The merged layer's concatenated output serves as the input for the CFFN. Ultimately, the output of CFFN is summed together at the final node, which gives the popularity score for the specific social media post. Mean Squared Error (MSE) is then calculated using both the ground truth and forecasted popularity scores. The computed error is backpropagated and weights are updated accordingly. 
For each iteration output vectors for the social feature, demographic feature, hashtag, and sentiment feature are calculated as illustrated in~\autoref{eq:ff1}, \autoref{eq:ff2}, \autoref{eq:ff3}, \autoref{eq:ff4}.
\begin{equation}
S_i = Conv1D_s3(Conv1D_s2(Conv1D_s1(f_i^{s})))
\label{eq:ff1}
\end{equation}

\begin{equation}
D_i = Conv1D_f3(Conv1D_f2(Conv1D_f1(f_i^{d})))
\label{eq:ff2}
\end{equation}

\begin{equation}
H_i = Conv1D_h3(Conv1D_h2(Conv1D_h1(f_i^{h})))
\label{eq:ff3}
\end{equation}

\begin{equation}
St_i = Conv1D_{st}3(Conv1D_{st}2(Conv1D_{st}1(f_i^{st})))
\label{eq:ff4}
\end{equation}
where $i=1,2,\hdots,N$. Here, $S_i$,$D_i, H_i$, and $St_i$ denote the social, demographic, hashtag, and sentiment feature vectors obtained after passing through CNN layers.
Further, we have flattened $S_i, D_i, H_i$, and $St_i$ and concatenated these feature vectors along with the hashtag-guided content feature vector ($\tilde{c}$) and denote the final merged vector as $M_i$ where 
\begin{equation}
M_i =[S_i, F_i, H_i, St_i, C_i]
\end{equation}
Here, $M_i$ denotes the concatenated feature vector which has a size of 27104. 
\subsection{Popularity Prediction}
\begin{figure}[!ht]
\includegraphics[width=\textwidth]{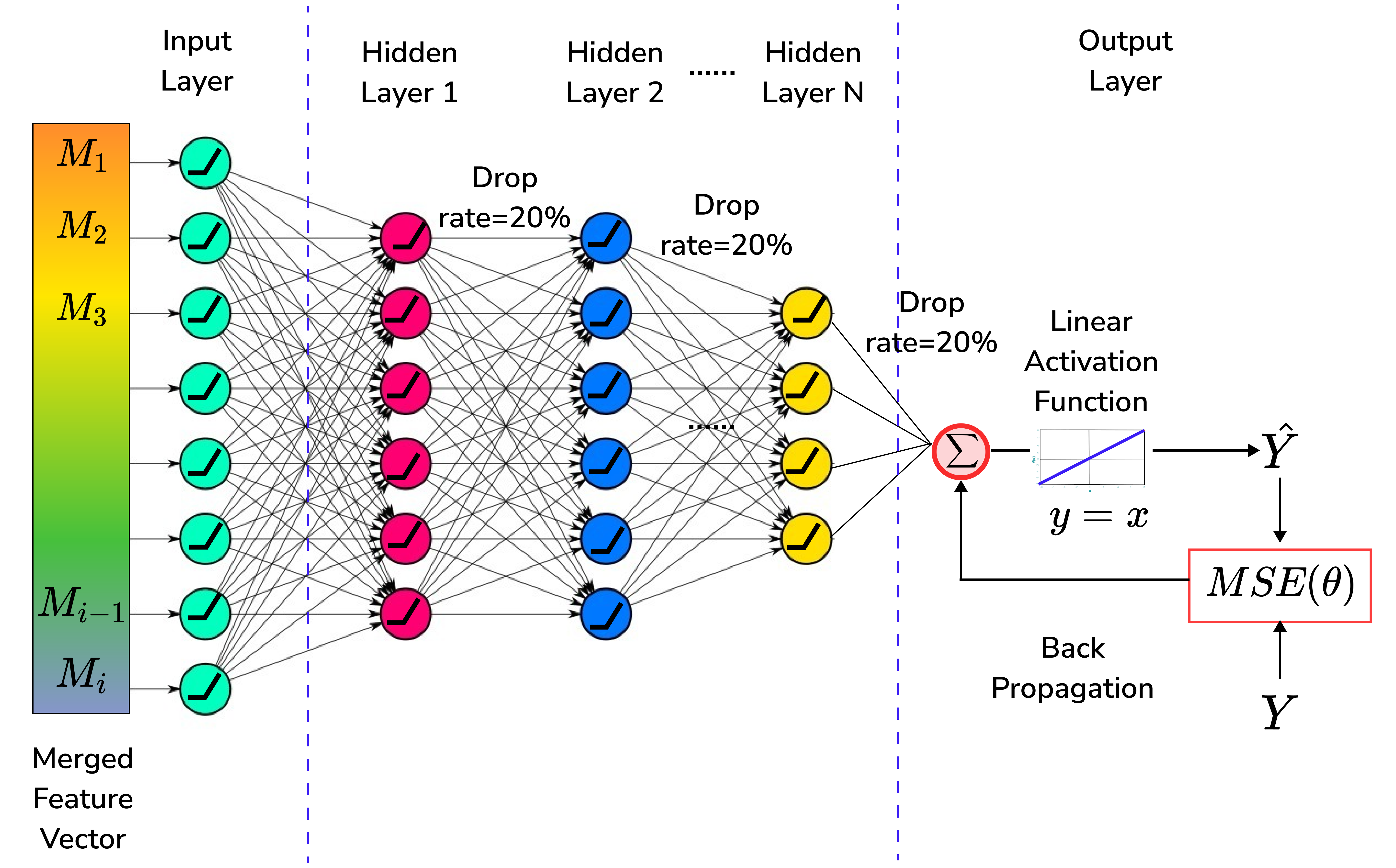}
\caption{Deep Feedforward Neural Network for Popularity Score Prediction}
\label{fig:pop_pred}
\end{figure}

We employ a Deep Feedforward Neural Network (DNN) as illustrated in \autoref{fig:pop_pred}, consisting of 12 fully connected layers (denoted by $N$) with decreasing sizes (13552, 6776, 3388, 1694, 847, 424, 212, 106, 53, 27, 13, 1) to forecast the popularity score. Each hidden layer is followed by a ReLU activation function and a dropout \cite{Srivastava2014} layer with a rate of 0.2. The output layer utilizes a linear activation function to directly predict the continuous star count.
\\
The final popularity score can be calculated as given in~\autoref{eq:ff5}.
\begin{equation}
 \hat{Y_i}= DNN(M_i)
 \label{eq:ff5}
\end{equation}
Here, $\hat{Y_i}$ and $Y_i$ are predicted and ground-truth popularity score of post $p_i$, and, DNN is the Deep Feedforward Neural Network.
The training objective is defined in~\autoref{eq:loss}.
\begin{equation}
MSE(\theta) = \min_{\theta} \left( \frac{1}{2|P|} \sum_{i=1}^|P| (\hat{Y_i} - Y_i)^2 \right)
\label{eq:loss}
\end{equation}
Here, $|P|$ is the number of posts in the training data and $\theta$ represent the NARRATOR's parameters. These are trained via back-propagation by maximizing the Mean Squared Error (MSE) cost function after being set with random values between -1 and 1.
\subsubsection{Theoretical Underpinnings of Feature Fusion}
The proposed model employs a multi-stage architecture for feature extraction and fusion:
\begin{itemize}
\item{Feature Retrieval:} Separate Convolutional Neural Networks (CNNs) are employed to retrieve relevant features from each modality (social context, demographics, hashtags, sentiment). This allows the model to learn modality-specific patterns.
\item{Dimensionality Reduction:} Techniques such as Principal Component Analysis (PCA) have been applied to social features to reduce redundancy and improve computational efficiency while preserving essential information.
\item{Fusion Network:} The retrieved features from each modality are then concatenated and fed into a Deep Feedforward Neural Network (DNN). This network learns to integrate the complementary information from different modalities, potentially revealing latent relationships that contribute to post popularity.
\end{itemize}
The theoretical foundation of this approach lies in the concept of multimodal representation learning. By learning a unified representation that captures the combined influence of different modalities, the model can potentially achieve superior prediction performance compared to models that rely on individual modalities.
\section{Experimental Evaluations}
\label{sec:exp_results}
Following a detailed account of the experimental setup, this section presents a comprehensive analysis of the data obtained from the experiments. This analysis will provide valuable insights into the effectiveness of our proposed approach.
\subsection{Experimental Setup}
This section details the datasets used in the study and the adopted preprocessing procedures. We next go through several comparison techniques, which are then followed by evaluation metrics.
\subsubsection{Datasets}
In this section, we cover various datasets on which experiments were conducted, followed by strategies for dataset preprocessing.
\begin{itemize}
\item SMP: The Social Media Prediction (SMP)~\citep{wu2019smp} is a real-world dataset provided by ACM Multimedia Grand Challenge in 2019. Initially, the raw data collection consisted of about 432K posts that were gathered from 135 distinct users' personal Flickr albums. In the dataset, each post has a unique picture ID that identifies the image and a user ID that identifies the person who uploaded it, the date the post was created, how many comments it received, how many hashtags were used, whether any users were tagged in the image, number of words in the title and the image caption. User-centric information such as the average view count, average member count, and group count were also collected as part of the data. Each image has a label that reflects its popularity based on the number of log-normalized views.
\item TPIC: TPIC2017~\citep{wu2017sequential} is a social media dataset for temporal popularity prediction that contains 680K photos and accompanying photo-sharing records on Flickr over a three-year period. The TPIC2017 dataset is diverse, containing photos, user data, and time information.
\end{itemize}
\paragraph{Dataset Preprocessing}
We preprocess the input data to convert it into an appropriate format to extract coherent features and accurately predict the popularity of multimodal posts. We apply several adaptations and normalization techniques to these datasets. To this end, we have taken all posts that contain hashtags, titles, and faces in the image. This left us with a total of 21,000 samples in the SMP dataset and 11,000 samples in the TPIC dataset. For our experiments, 80\% of the posts were utilized for training, 10\% for validation, and 10\% for testing.
\subsubsection{Compared Methods}
In this part, we discuss various state-of-the-art methods for assessing the effectiveness of the suggested framework.
\begin{enumerate}
\item{Leveraging Hashtag Networks for Multimodal Popularity Prediction of Instagram Posts (HashPop): Liao \textit{et al.}~\citep{liao-2022-leveraging}} has predominantly used hashtags as a separate modality to gauge the popularity of a specific post. The authors have constructed a graph-like hashtag network where hashtags used in the post serve as nodes of the graph. Edges are created between hashtags if many hashtags appear in a single post. This combined vector is then represented as a hashtag feature. In addition, authors used InceptionV3 to obtain image embedding, and sentence transformer to obtain caption embedding for the post along with additional metadata. The overall post-representation is obtained by concatenating these obtained features. The output from the last layer is utilized to calculate the popularity score in a dense layer once this composite representation has been passed through.
\item{Multimodal Deep Learning Framework for Image
Popularity Prediction on Social Media (VSCNN): Abousaleh \textit{et al.}~\citep{ abousaleh2020multimodal}} chose social metadata, post metadata, and time metadata as primary components for predicting post popularity. Low-level, high-level, and Deep Learning-based visual attributes are extracted from the image that constitutes a given post. Furthermore, after feature extraction, the authors employed CNNs on visual and social metadata features independently to learn their high-level representations. At the output layer, shared multimodal properties are learned and the popularity score is determined. To this end, the output of two separate CNN networks is then integrated into a shared network which is made up of one merge layer and two dense layers to predict the popularity of a post.
\item{Multimodal Deep Learning for Social Media Popularity Prediction With Attention Mechanism (MMAtt): Xu \textit{et al.}~\citep{xu2020multimodal}} provided an attention mechanism for forecasting post popularity. The authors retrieved four distinct features namely categorical, numerical, visual, and textual. The categorical feature consists of category, subcategory, and path alias of post embeddings which were obtained and passed to individual dense layers, the output of which was further passed to the common dense layer. The numerical information contains the number of followers, the date the post was created, and the location. The authors applied ResNet50 to extract visual features which were later sent to a dense layer. The textual features contain the title and hashtags of the post, and embeddings of both were obtained using word2vec and were passed to two-layer LSTM and individual dense layers. The output of all dense layers was concatenated and transferred to a single dense layer, after which an attention layer was utilized to compute a popularity score.
\item{Social Media Popularity Prediction: A Multiple Feature Fusion Approach with Deep Neural Networks (FuseDNN): Ding \textit{et al.} \citep{ding2019social}} retrieved visual, textual, user, temporal, and geographical location features. In visual features, the authors retrieved ResNet-101 characteristics, an intrinsic popularity score, and an aesthetics score from the post's provided image. Tag count, title length, and title embedding using BERT are all included in text features. While the time component contains the post duration and the location information provides geographic coordinates, user data includes the count of followers, followings, and posts. The remaining entities were concatenated and passed to a different dense layer. The ResNet-101 and BERT embeddings were passed to separate dense layers. To forecast the popularity score, the output from all three dense layers was combined and transferred to three dense layers to yield the predicted popularity score.
\item{Predicting Tweet Engagement with Graph Neural Networks (TweetGage): Arazzi \textit{et al.} \citep{arazzi2023predicting}} developed a novel Graph Neural Network framework for predicting user engagement on social media. TweetGage leverages a graph-based model based on hashtag relationships, capturing semantic connections beyond individual post features.  
\item{Gradient Boost Tree Network based on Extensive Feature Analysis for Popularity Prediction of Social Posts (MFTM): Hsu \textit{et al.}~\citep{hsu2023gradient}} presented a multi-modality feature mining framework for social media popularity prediction. It leverages identity-related user features alongside traditional modalities (text and image) to achieve significant performance improvements, suggesting a stronger influence of identity compared to other user metadata. LightGBM and TabNet are employed to capture complex relationships within the enriched feature set.
\item{Enhanced CatBoost with Stacking Features for Social Media
Prediction (ECSF): Mao \textit{et al.}~\citep{mao2023enhanced}} proposed a novel social post popularity prediction approach utilizing enriched post and user features. ECSF employs innovative stacking features to 
capture higher-order interactions between text and image features, potentially leading to a more comprehensive understanding of the underlying factors that influence social media post popularity. ECSF's effectiveness is substantiated by its state-of-the-art performance on the SMP challenge dataset, surpassing prior methods that primarily relied on extracting lower-order features.
\end{enumerate}
\subsubsection{Evaluation Metrics}
In this study, we adopted Mean Squared Error (MSE) and Mean Absolute Error (MAE) as the primary metrics to quantify the prediction accuracy of our model.
\begin{itemize}
\item MSE: To determine the mean of the squared sum of prediction errors, MSE is frequently used. Each prediction error represents the discrepancy between a data point's actual value and the value estimated by a regression model. It is simpler to determine the gradient of MSE since it has straightforward mathematical features. Due to its computational simplicity, smooth differentiability, and greater optimization amenability, MSE is typically supplied as the default measure for the majority of predictive models. A serious flaw with MSE is that it squares large prediction errors, which strongly penalizes them. Due to the quadratic accumulation of each MSE error, the overall error is significantly influenced by outliers in data. This demonstrates that MSE undervalues the model's performance because of its high sensitivity to outliers and the disproportionate weight assigned to their effects, MSE undervalues the model's performance in this case. When outliers are present in the data, only then the disadvantage of MSE becomes obvious, making MAE an adequate replacement. MSE is defined as given in~\autoref{eq:eq16}.
\begin{equation}
    MSE=\frac{1}{\lvert N\rvert}\\\sum_{i=1}^N(y_i-\hat{y}_i)^2 
     \label{eq:eq16}
\end{equation}
Here, $N$ denotes the total number of posts, $\hat{y}_i$ denotes the predicted popularity score, and $y_i$ denotes the actual popularity score of the $i^{th}$ multimodal post ($p_i$).
\item MAE: MAE is a straightforward metric typically used to assess the precision of a regression model. It calculates the mean absolute value of each prediction mistake made by the model overall test set samples. Unlike metrics sensitive to outliers, MAE assigns equal weight to all prediction errors. This characteristic ensures that larger deviations from the actual values contribute linearly to the overall error score. However, MAE focuses solely on the absolute magnitude of the error, without considering the direction of the discrepancy (overprediction or underprediction). Consequently, MAE provides an objective measure of the model's overall performance in terms of absolute prediction error, but it doesn't necessarily indicate whether the model consistently overestimates or underestimates the target variable. By employing the absolute value of the prediction error rather than its squared value, MAE becomes more resilient to outliers than MSE because it does not penalize significant errors as heavily as MSE. Therefore, MAE has both benefits and drawbacks. While it helps in managing outliers, it does not penalize significant forecasting errors. MAE is described in~\autoref{eq:mae}.
\end{itemize}
\begin{equation}
    MAE=\frac{1}{\lvert N\rvert}\\\sum_{i=1}^N{\lvert y_i-\hat{y}_i \rvert} 
     \label{eq:mae}
\end{equation}
Here, $N$ denotes the total number of posts, $\hat{y}_i$ denotes the predicted popularity score, and $y_i$ denotes the actual popularity score of the $i^{th}$ multimodal post $(p_i)$.
\subsection{Experimental Results}
The experimental results for the proposed and current state-of-the-art approaches on various datasets, ablation investigations, visualization of predictions from both existing and proposed methods, statistical analysis, ranking of important features, implementation details, and limitations are all covered in this section.
\subsubsection{Effectiveness Comparison}
\begin{figure}[H]
\includegraphics[clip,width=\textwidth]{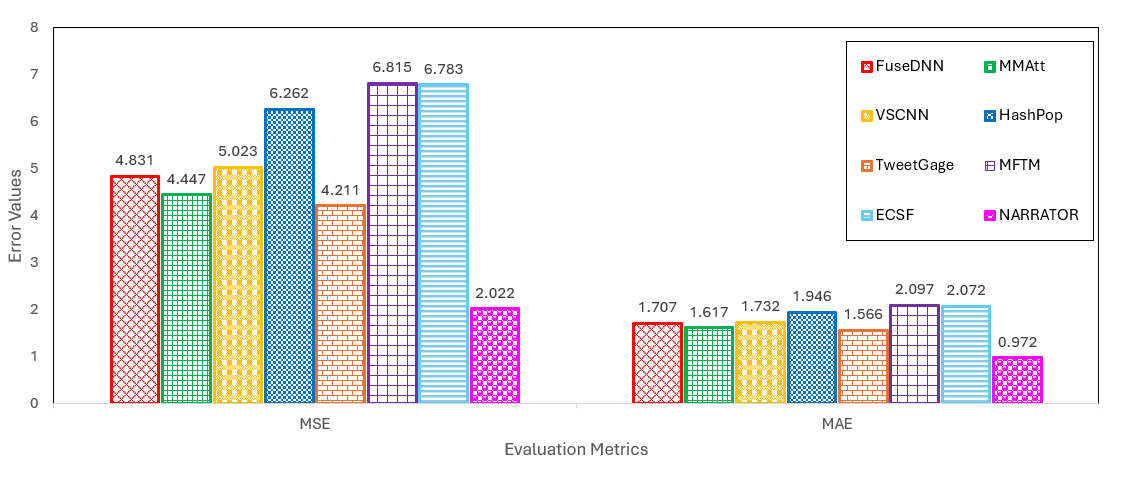}
\caption{Performance Comparison Curves on SMP Dataset}
\label{fig:smp1}
\end{figure}

To assess the performance of our suggested technique on various datasets, we compare it to the existing methods on TPIC and SMP datasets. 
\begin{table}[!ht]
\centering
\caption{Effectiveness Comparison Results on Different Datasets}
\label{table:res1}   
    \begin{tabular*}{0.9\linewidth}{@{\extracolsep{\fill}} lcccc }
    \toprule
        \textbf{Methods} & \multicolumn{2}{c}{\textbf{TPIC}} & \multicolumn{2}{c}{\textbf{SMP}} \\ 
        \cmidrule(l){2-3} \cmidrule(l){4-5}
            &\textbf{MSE}&\textbf{MAE}
            &\textbf{MSE}& \textbf{MAE}\\         
    \midrule
        {FuseDNN, Ding \textit{et al.}'19}& 2.716 & 1.318 & 4.831 & 1.707\\
        {MMAtt, Xu \textit{et al.}'20} & 2.367 & 1.170 & 4.447 & 1.617\\
        VSCNN, Abousaleh \textit{et al.}'20 & 1.711 &1.015 & 5.023 & 1.732\\
        {HashPop, Liao \textit{et al.}'22} & {3.200}& {1.435} & 6.262 & 1.946\\
        {TweetGage, Arazzi\textit{et al.}'23} & 2.132 & 1.145 & 4.211 & 1.566\\
        {MFTM, Hsu \textit{et al.}'23} & 5.264 & 1.946 & 6.815 & 2.097 \\
        ECSF, Mao \textit{et al.}'23& 5.115 & 1.911 & 6.783 & 2.072 \\
        \textbf{NARRATOR}   &\textbf{1.196} &\textbf{0.854} 
        &\textbf{2.022} &\textbf{0.972} \\
    \bottomrule    
    \multicolumn{5}{@{}l}{\footnotesize MSE: Mean Squared Error, \footnotesize MAE: Mean Average Error}    
    \end{tabular*}   
\end{table}
The performance comparison using the aforementioned datasets is given in~\autoref{table:res1}. It is evident from~\autoref{table:res1} that our suggested model, NARRATOR, performs noticeably better than state-of-the-art methods. NARRATOR beats HashPop with a 62.625\% improvement in MSE and 40.487\% in MAE on the TPIC dataset and 67.709\% improvement in MSE and 50.051\% in MAE on the SMP dataset. The authors in HashPop have focused more on using the hashtag network besides content-based and metadata information. Our proposed model which consists of demographic and sentiment information embedded in captions and hashtags along with hashtag-guided attention content features beats HashPop. NARRATOR beats VSCNN with a relative improvement of 30.099\% in terms of MSE and 15.862\% in terms of MAE on the TPIC dataset and 59.745\% in MSE and 78.189\% in MAE on the SMP Dataset. VSCNN predicted popularity based on social and visual features whereas in our model we have incorporated four additional features namely demographic information, hashtag, textual, sentiment of caption, and hashtags besides a hashtag-guided attention mechanism employed on content-based features. NARRATOR beats MMAtt with a relative improvement of 49.471\% in MSE and 27.008\% in MAE on the TPIC dataset and 54.531\% in MSE and 39.888\% in MAE on the SMP dataset. While the authors incorporated an attention mechanism within their model, NARRATOR achieves demonstrably stronger performance. This improvement can be attributed to our implementation of a hashtag-guided attention mechanism that models the influence of hashtags on linguistic and visual features. Our attention mechanism produces superior results because the hashtags are very closely related to the title and the image associated with the post, which in turn boosts the model's performance significantly. The proposed model beats FuseDNN exhibiting a relative improvement of 55.964\%  and 35.204\%  in MSE and MAE on the TPIC dataset and 58.145\% improvement in MSE and 43.057\% in MAE on the SMP dataset. NARRATOR surpasses TweetGage exhibiting a relative improvement of 43.092\%  and 25.415\%  in terms of MSE and MAE on the TPIC dataset and 51.982\% improvement in terms of MSE and 37.931\% in terms of MAE on the SMP dataset. Unlike TweetGage which captures relationships among posts based on common hashtags, we employ the topical, structural, semantic, and sentiment information from hashtags besides image, caption, and demographics. The proposed model beats MFTM exhibiting a relative improvement of 77.279\%  and 56.115\%  in terms of MSE and MAE on the TPIC dataset and 70.330\% improvement in terms of MSE and 53.648\% in terms of MAE on the SMP dataset. After feature extraction, MFTM employs an ensemble of TabNet and LightGBM which are Machine Learning models to predict post popularity whereas NARRATOR employs a hashtag-guided attention mechanism on content features and a a deep neural network to forecast the post popularity. NARRATOR surpasses ECSF by a relative improvement of 76.617\% in MSE  and 55.311\%  in MAE on the TPIC dataset and 70.190\% improvement in MSE and 53.088\% in MAE on the SMP dataset. ECSF relies on feature stacking and a CatBoost model for prediction, which limits its ability to capture complex relationships between features. NARRATOR's deep neural network with a hashtag-guided attention mechanism overcomes this limitation by learning more intricate feature interactions.
\begin{figure}[H]
\includegraphics[clip,width=\textwidth, height=5.41cm]{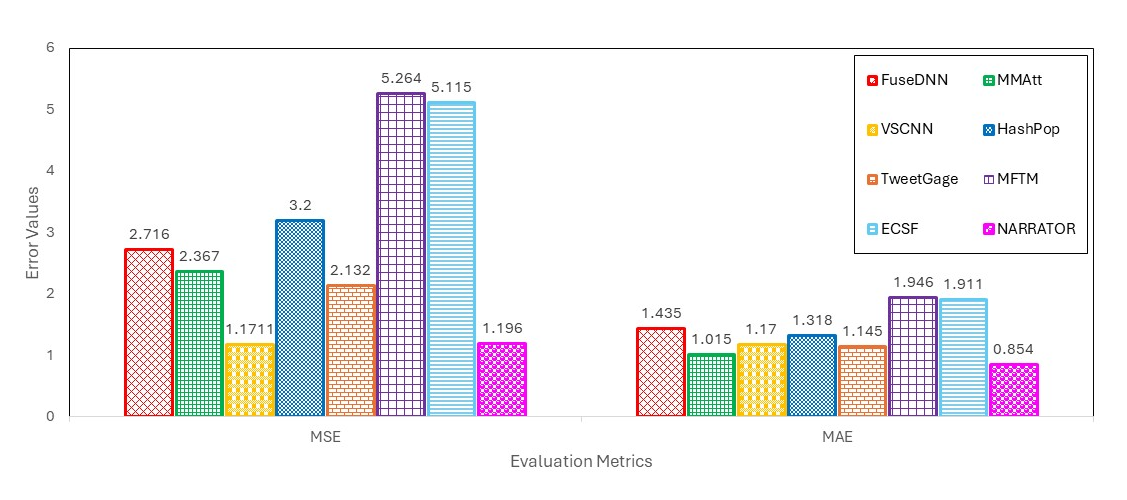}
\caption{Effectiveness Comparison Curves on TPIC Dataset}
\label{fig:tpic1}
\end{figure}
As illustrated in \autoref{fig:tpic1} and~\autoref{fig:smp1}, NARRATOR demonstrates superior performance compared to all four baseline models on both datasets. This is evident in the consistently lower MSE and MAE values achieved by NARRATOR. Notably, on the TPIC dataset, NARRATOR achieves the lowest MSE of 1.196 and the lowest MAE of 0.854. Similarly, on the SMP dataset, NARRATOR outperforms other models with an MSE of 2.022 and an MAE of 0.972. These results suggest that NARRATOR effectively captures the underlying factors influencing social media post popularity across different datasets.
This implies that our derived features, such as sentiments, demographics, and hashtags, are essential for determining how popular social media posts are. The fact that hashtags influence both textual and visual content features and enrich the overall representation of the posts in predicting post popularity on social network platforms is another important result. 
\subsubsection{Ablation Studies}
In this section, we discuss the model's performance by analyzing feature combinations, and the effectiveness of different attention mechanisms on the performance of the proposed model.
\paragraph{Feature Ablation}
We investigate the role of our two novel features- visual demographics and sentiment extracted from hashtags in enhancing popularity prediction. 
\begin{table}[!ht]\footnotesize
\centering
\caption{Feature Ablation Study}
\label{table:res4}   
    \begin{tabular*}{0.9\linewidth}{@{\extracolsep{\fill}} lcccc }
    \toprule
        \textbf{Variant} & \multicolumn{2}{c}{\textbf{TPIC}} & \multicolumn{2}{c}{\textbf{SMP}} \\ 
        \cmidrule(l){2-3} \cmidrule(l){4-5}
            &\textbf{MSE}&\textbf{MAE}
            &\textbf{MSE}& \textbf{MAE}\\   \midrule
NARRATOR w/o (Sentiment from Hashtags+ Demographics)
& 1.387 & 0.912 & 2.367 & 1.084\\
NARRATOR w/o Sentiment from Hashtags
& 1.372 & 0.872 & 2.346 & 1.092\\
NARRATOR w/o Demographics & 1.238 & 0.897 & 2.465 & 1.131\\
        \textbf{NARRATOR} &\textbf{1.196} &\textbf{0.854} &\textbf{2.022} &\textbf{0.972} \\
    \bottomrule    
    \multicolumn{5}{@{}l}{\footnotesize MSE: Mean Squared Error, \footnotesize MAE: Mean Average Error}    
    \end{tabular*}   
\end{table}
\autoref{table:res4} demonstrates that the complete NARRATOR model, incorporating both novel features- visual demographics and sentiment extracted from hashtags, achieves superior performance compared to ablated versions. The complete model achieves the lowest MSE and MAE values across both TPIC and SMP datasets, highlighting the significant contribution of these features to accurate popularity prediction. This degradation is particularly pronounced when both features are removed simultaneously, resulting in an absolute increase of 13.77\% and 6.36\% in MSE and MAE on TPIC, and 14.58\% and 10.33\% on SMP, respectively. Removing demographics and sentiment features from hashtags affects the model's ability to understand the target audience and the emotional tone of the post, both of which could be relevant for popularity. The exclusion of visual demographics leads to an absolute increase of 3.39\%, 4.79\% in MSE and MAE on TPIC and 17.97\% and 14.06\% in MSE and MAE on SMP dataset, respectively. This highlights the importance of understanding the target audience. Visual demographics provide insights into the age group, gender, and other visual cues that may resonate with specific user segments, enabling the model to better predict post appeal. The exclusion of sentiments from hashtags leads to an absolute increase of 12.83\%, 2.06\% in MSE and MAE on TPIC and 13.81\% and 10.99\% in MSE and MAE on SMP dataset, respectively. This emphasizes the role of emotional tone in driving post popularity. Hashtags often encapsulate the sentiment or theme associated with a post. By incorporating hashtag sentiment, the model can gauge the emotional appeal of the content, which is a key factor influencing user engagement and sharing behavior. In essence, visual demographics help the model understand who the content might appeal to, while hashtag sentiment helps understand how the content might make the audience feel. By integrating both, NARRATOR gains a more comprehensive understanding of the factors driving popularity, enabling it to make more accurate predictions. This underscores the synergistic effect of combining visual demographics and hashtag sentiment for understanding and predicting multimodal post popularity.
\paragraph{Attention Mechanisms}
To demonstrate the significance of the novel hashtag-guided attention mechanism, we compare its performance with various attention mechanisms discussed below. 
 \begin{table}[!h]
\centering
\caption{Effectiveness of Attention Mechanisms}
\label{table:att}   
    \begin{tabular*}{0.9\linewidth}{@{\extracolsep{\fill}} lcccc }
    \toprule
        \textbf{Attention Technique} & \multicolumn{2}{c}{\textbf{TPIC}} & \multicolumn{2}{c}{\textbf{SMP}} \\ 
        \cmidrule(l){2-3} \cmidrule(l){4-5}
            &\textbf{MSE}&\textbf{MAE}
            &\textbf{MSE}& \textbf{MAE}\\ \midrule
        {$NARRATOR_{NA}$} & 1.510 & {0.917} &{3.715}& {1.417}\\
        $NARRATOR_{SA}$ & 1.567 & 0.944 & 3.474 & 1.445\\
        $NARRATOR_{CA}$ & 1.486 & 0.916 & 2.904 & 1.218\\
        $NARRATOR_{PA}$ & 1.389 & 0.905 & 3.367 & 1.359\\
        \textbf{$NARRATOR_{HGA}$} & \textbf{1.196} & \textbf{0.854} & \textbf{2.022} & \textbf{0.972} \\
    \bottomrule    
    \multicolumn{5}{@{}l}{\footnotesize MSE: Mean Squared Error, \footnotesize MAE: Mean Average Error}    
    \end{tabular*}   
\end{table}
Table \ref{table:att} showcases the performance of NARRATOR with different attention mechanisms. Here, the variants that use no attention, self-attention, cross-attention, parallel co-attention, and hashtag-guided attention are $NARRATOR_{NA}$, $NARRATOR_{SA}$, $NARRATOR_{CA}$, $NARRATOR_{PA}$, $NARRATOR_{HGA}$ respectively. 
The performance difference when NARRATOR is implemented without any attention mechanism is 23.68\% and 9.53\% in MSE and MAE on the TPIC dataset and 41.80\% and 32.73\% on the SMP dataset compared to hashtag-guided attention. The performance of NARRATOR is the lowest in the absence of any attention mechanism. Compared to self-attention, hashtag-guided attention shows and absolute improvement of 23.68\%, 9.53\% in MSE and MAE on TPIC dataset and 41.80\% and 32.73\% on SMP dataset. The superior performance of hashtag-guided attention over self-attention indicates that solely modeling intra-modal relationships (within text or image) is less effective than incorporating the semantic context provided by hashtags. Compared to cross-attention, hashtag-guided attention shows and absolute improvement of 19.52\%, 6.77\% in MSE and MAE on TPIC dataset and 30.37\% and 20.20\% on SMP dataset. While cross-attention improves performance by modeling inter-modal interactions (between text and image), hashtag-guided attention further refines this by leveraging the contextual cues embedded in hashtags, leading to a better understanding of content. Our hashtag-guided attention mechanism outperforms parallel co-attention, demonstrating an absolute improvement of 13.89\% and 56.35\% on TPIC and 39.95\% and 39.81\% on SMP in terms of MSE and MAE metrics, respectively. Unlike the parallel co-attention mechanism that solely focuses on the relationship between text and image features, the proposed hashtag-guided attention mechanism introduces a crucial element: the influence of hashtags on both modalities. The superior performance of hashtag-guided attention compared to parallel co-attention emphasizes the value of explicitly incorporating hashtag semantics into the attention mechanism. Hashtags provide a bridge between textual and visual content, allowing the model to focus on the most relevant aspects of both modalities for popularity prediction.
\paragraph{Advantages of Hashtag-Guided Attention}
Hashtag-guided attention utilizes hashtag embeddings to guide the model's focus towards features in both text and images that align semantically with the hashtags. This targeted approach allows the model to:
\begin{enumerate}
\item Grasp the intended meaning and purpose of the post, as hashtags often encapsulate the core message.
\item Bridge contextual gaps, going beyond literal content to interpret nuances that contribute to popularity.
\end{enumerate}
By combining content analysis with hashtag-driven context, our approach achieves a more holistic understanding of social media posts, leading to more accurate popularity predictions.
\subsubsection{Qualitative Analysis}
Predicting the popularity score of the given post is a common evaluation protocol in popularity prediction tasks. This section presents a qualitative evaluation aimed at understanding how accurately our proposed model predicts the popularity of social media posts. Example posts with accompanying captions, images, and associated hashtags are illustrated. We also display the ground-truth popularity scores and popularity scores predicted by state-of-the-art methods. These example posts have been chosen randomly from test data. 
\begin{figure*}[!ht]
\subfloat[Post 1]{%
\includegraphics[width=0.5\textwidth]{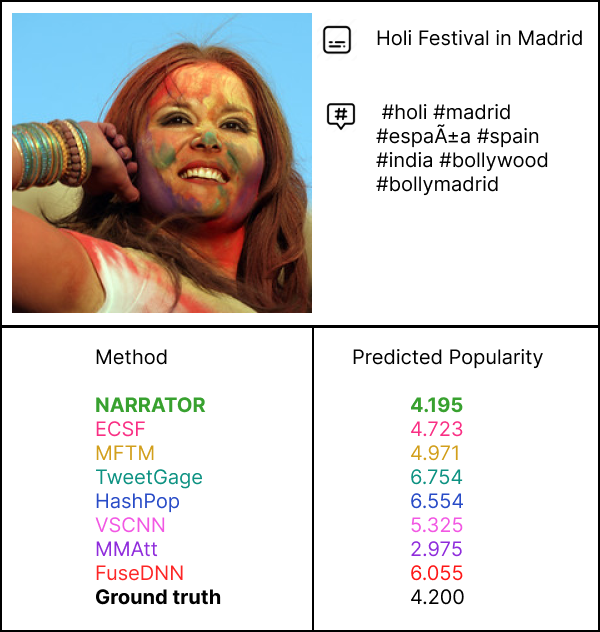}\label{fig:qa1}%
}
\subfloat[Post 2]{%
\includegraphics[width=0.5\textwidth]{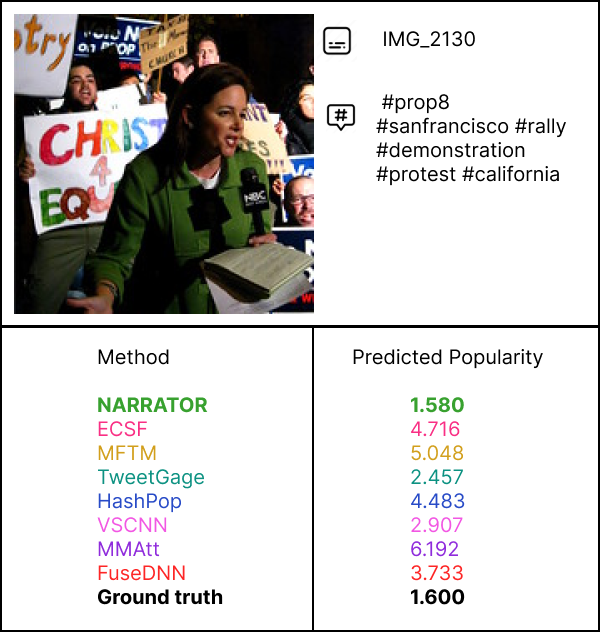}\label{fig:qa2}%
}
\caption{Posts Depicting Popularity Scores Predicted by Different Methods}
\label{fig:figres1}
\end{figure*}
As can be seen in~\autoref{fig:qa1}, our proposed model predicts a popularity score that is very close to the ground-truth popularity score. The caption of the first example post i.e., ``Holi festival in Madrid", bears a striking resemblance to the accompanying image, which depicts a girl celebrating the festival. Hashtags such as \#holi and \#spain indicate a cultural celebration happening outside its traditional location (India), \#india and \#bollywood suggest the post might target the Indian diaspora or Bollywood fans in Spain. The model focuses on visual features in the image related to the Holi celebration (colors, people celebrating). Textual features in the caption (``Holi festival") are analyzed alongside hashtags such as \#india to understand the cultural significance. Hashtags such as \#spain and \#bollymadrid help identify a potential audience interested in Indian culture or Bollywood within Spain. By considering these contextual cues from hashtags, the model refines its understanding of the post's content and target audience. This validates our hypothesis that modeling the interaction between the title and image under the influence of hashtags via a hashtag-guided attention mechanism effectively captured the relevant aspects for predicting engagement and considerably enhanced the model's performance. Furthermore, our experimental results show that sentiments are related to post popularity. The hashtags \#holi and \#india are directly related to the post caption as Holi is the festival of colors that celebrates spring and the triumph of good over evil in India and is associated with joy, love, and new beginnings. Hashtag ``\#bollymadrid" combines Bollywood with Madrid, suggesting a fusion of Indian and Spanish culture, which can be seen as positive. Overall, the post promotes a positive and inclusive sentiment about cultural exchange. The positive mood communicated by captions and hashtags aids the model in anticipating better outcomes. Further, as illustrated in Figure~\autoref{fig:qa2}, the ground-truth popularity score for the post is 1.6. Our model's prediction closely approximates this value, achieving a score of 1.580. Overall, this research helps to visualize our model's ability to estimate post popularity accurately by leveraging innovative features.
\subsubsection{Statistical Analysis}
This section explores the statistical relationship between NARRATOR's predictions and real-world content popularity using two established correlation coefficients: Spearman's Rank Correlation Coefficient (SRCC) and Pearson Correlation Coefficient (PCC). To evaluate the NARRATOR's effectiveness, we calculated correlation coefficients based on predictions by NARRATOR and the ground-truth popularity scores of posts in TPIC and SMP datasets.
\begin{figure}[htbp]
    \centering
    \begin{subfigure}[b]{0.4\textwidth}
        \centering
        \includegraphics[width=\textwidth]{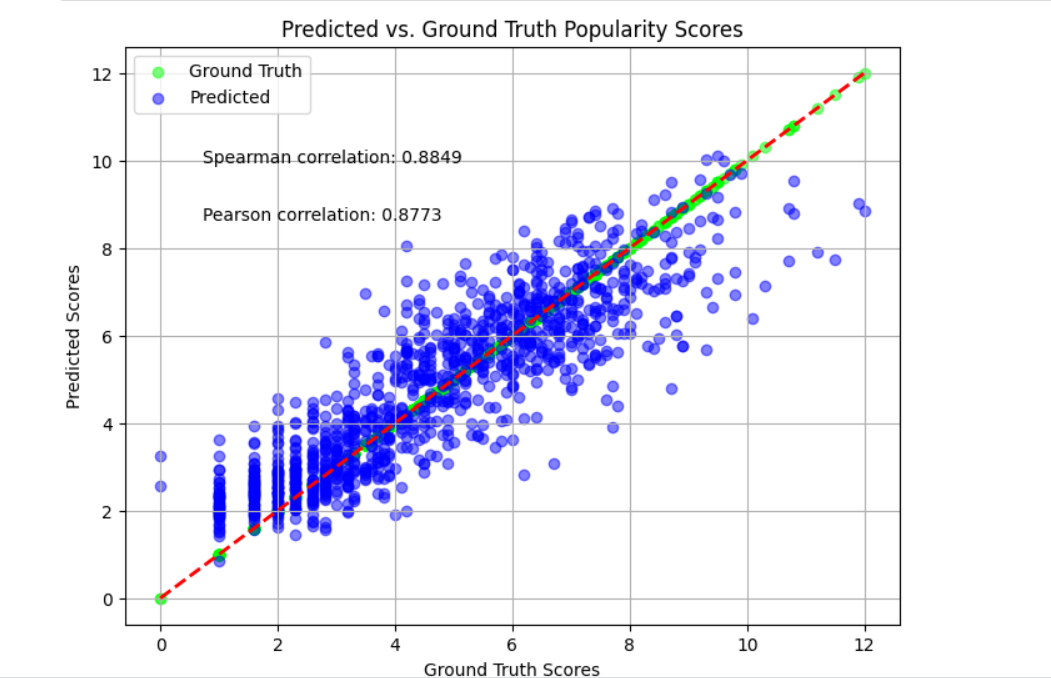}
        \caption{Correlation on TPIC Dataset}
        \label{fig:figure1}
    \end{subfigure}
    \hspace{2cm} 
    \begin{subfigure}[b]{0.4\textwidth}
        \centering
        \includegraphics[width=\textwidth]{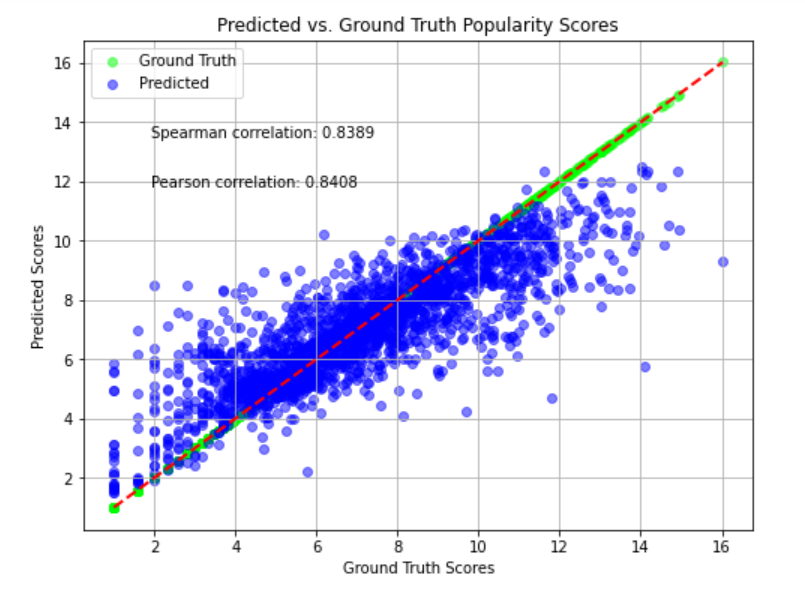}
        \caption{Correlation on SMP Dataset}
        \label{fig:figure2}
    \end{subfigure}
    \caption{Statistical Analysis}
    \label{fig:stats}
\end{figure}
Our analysis yielded significant correlations, visually depicted in the scatterplots of \autoref{fig:stats}. Here, proximity to a straight diagonal line signifies a robust correlation (shown in red).
As can be seen from \autoref{fig:stats}, the SRCC values were found to be 0.885 and 0.8773, while the PCC values were computed as 0.8355 and 0.8409 for TPIC and SMP datasets, respectively. 
Moreover, the calculated p-values for these correlations were determined to be highly significant ($p \leq 0.05$), affirming the robustness of the observed relationships between predicted and actual popularity levels across social media platforms. In summary, our research demonstrates the efficacy of NARRATOR in accurately forecasting the popularity of multimodal content, as evidenced by strong correlations established through rigorous statistical analyses conducted on real-world TPIC and SMP datasets. These findings have implications for various fields, including content marketing, audience engagement strategies, and social media analytics.
\subsubsection{Feature Correlation Analysis}
To investigate the influence of individual features on predicting post popularity, we calculated Spearman's Rank Correlation Coefficient (Spearman’s Rho) \cite{Spearman1904} with the popularity score.
\begin{table}[!ht]
\centering
\caption{Spearman's Rank Correlation of Features with Popularity Scores}
\label{tab:smp_correlation}
\begin{tabular}{lll}
\toprule
\textbf{Feature}& \textbf{SRCC (TPIC)} &\textbf{SRCC (SMP)} \\
\midrule
Post Id  & -0.529 & -0.244\\
User Id & -0.097 & -0.319\\
Title length  & 0.3999 & 0.018 \\
Description Length & 0.552 & -\\
Tagcount & 0.458 & 0.446\\
Avgview & 0.778 & -\\
Groupcount & 0.473 & -\\
Avgmembercount & 0.319 & -\\
Postday & -0.066 & 0.042\\
Posttime & 0.263 & -0.008\\
Postmonth & -0.112 & -0.074\\
Hashtags & 0.402 & 0.237 \\
Sentiment & 0.429 & 0.117\\
Demographic Features &  0.225 & 0.095\\
\bottomrule
\end{tabular}
\end{table}
The results for the TPIC and SMP datasets are shown in
Table~\ref{tab:smp_correlation}. Our derived features i.e., sentiment, hashtags, and demographic features exhibit a positive correlation on both datasets. On the TPIC dataset, the post id has the highest negative correlation and average views (avgview) has the highest positive correlation, whereas the user id has the highest negative correlation, and tag count has the highest positive correlation on the SMP dataset. However, correlation does not imply causation, and it might miss non-linear relationships. The drawback  Spearman's rank correlation coefficient can only detect monotonic relationships and is not suitable for non-linear relationships.
\subsubsection{Feature Ranking and Importance}
To assess the importance of derived features, we 
independently ranked the performance of each feature in terms of Mean Squared Error (MSE) and the Mean Absolute Error (MAE), with a higher rank of 1 indicating optimal performance. Following that, by aggregating the preliminary ranks acquired for each parameter, the resulting average rank was used as a comprehensive assessment of overall performance.
 \begin{table}[!ht]
\centering
\caption{Feature Ranking and Importance}
\label{table:feimp}   
    \begin{tabular*}{0.9\linewidth}{@{\extracolsep{\fill}} lcccc }
    \toprule
        \textbf{Feature} & \multicolumn{2}{c}{\textbf{TPIC}} & \multicolumn{2}{c}{\textbf{SMP}} \\ 
        \cmidrule(l){2-3} \cmidrule(l){4-5}
            &\textbf{MSE}/\textbf{MAE} &\textbf{Rank}
            &\textbf{MSE}/ \textbf{MAE}&\textbf{Rank}\\       
    \midrule
        {w/o Sentiment (Text)} & {$1.362_{(5)}$}/{$0.867_{(6)}$} & 5.5 & {$2.367_{(5)}$}/ {$1.084_{(6)}$} &5.5\\
        {w/o Sentiment (Hashtags)} &4.5{$1.372_{(4)}$}/{$0.872_{(5)}$} &4.5 & {$2.346_{(6)}$}/ {$1.092_{(5)}$} &5.5 \\
        w/o Hashtags & $2.553_{(1)}$ / $1.251_{(1)}$ &1 & $2.458_{(4)}$ / $1.125_{(4)}$ & 4\\
        w/o Demographics & $1.238_{(6)}$ / $0.897_{(4)}$ & 5 & $ 2.465_{(3)}$ /$1.131_{(3)}$ &3 \\
        w/o Social & $2.166_{(2)}$ / $1.093_{(2)}$ &  2& $2.897_{(2)}$/$ 1.267_{(1)}$& 1.5\\
        w/o Content & $1.435_{(3)}$ / $0.908_{(3)}$ &  3& $3.021_{(1)}$/$1.228_{(2)}$& 1.5\\
    \bottomrule    
    \multicolumn{5}{@{}l}{\footnotesize MSE: Mean Squared Error, \footnotesize MAE: Mean Average Error}    
    \end{tabular*}   
\end{table}
~\autoref{table:feimp} highlights the significance of both established and newly introduced features in predicting post popularity. While certain features such as social information and hashtags consistently rank high across both datasets, showcasing their fundamental role in capturing engagement and contextual information, respectively, the impact of other features such as sentiment and visual demographics appears to be more nuanced and context-dependent. The relatively lower individual ranks of the newly introduced features (sentiments from hashtags and visual demographics) might not fully reflect their true value. Their strength lies in their synergistic contribution to the overall model, capturing subtle nuances and previously overlooked aspects of popularity dynamics. In particular, they address specific gaps in prior research, offering a more comprehensive understanding of the factors influencing post popularity. Furthermore, the importance of features can vary depending on the specific platform and its user base. While some features might be universally influential, others might play a more significant role in specific contexts or for particular types of posts. Our analysis underscores the complex and multifaceted nature of popularity prediction, highlighting the need for a holistic approach that considers a diverse range of features and their interactions.
\subsubsection{Implementation Details}
The experiments leveraged a high-performance computing environment featuring a Linux server architecture. The server's processing power included an Intel(R) Xeon(R) Silver 4215R CPU operating at 3.20 GHz, complemented by 256 GB of RAM and a dedicated NVIDIA Tesla T4 GPU with 16 GB of memory. This configuration facilitated efficient model training and experimentation. To achieve optimal model performance, a rigorous hyperparameter tuning process was conducted. The learning rate was meticulously set to 0.0001, ensuring convergence without excessive learning speed. A batch size of 20 was chosen to strike a balance between computational efficiency and gradient estimation accuracy. The Adam optimizer, renowned for its adaptive learning rate capabilities, was employed to facilitate efficient optimization. Training proceeded for a maximum of 30 epochs, incorporating an early stopping mechanism with a patience of 5 epochs to prevent overfitting. Additionally, a dropout rate of 0.2 was strategically applied after each layer within the model architecture to further mitigate overfitting tendencies. 
To ensure a level playing field for performance evaluation, the embedding dimension (D) was consistently set to 768 for all comparative methods employed in the study.
Before feeding image data into the VGG-19 network, all image samples underwent a standardized pre-processing step. This step involved rescaling each image to a uniform size of 224 x 224 pixels. This normalization ensured consistent image representation and facilitated network training. In Deep Feed Forward Network, we employ a series of  12 fully connected layers of size 13552, 6776, and so on till size 1.
\subsubsection{Computational Cost Analysis}
To assess the cost implications of our proposed model, NARRATOR, we investigated its runtime and scalability. 
\begin{table}[!ht] \footnotesize
\centering
\caption{Computational Time Analysis}
\label{table:res_time}   
    \begin{tabular*}{0.9\linewidth}{@{\extracolsep{\fill}} lcccc }
    \toprule
        \textbf{Methods} & \multicolumn{2}{c}{\textbf{TPIC}} & \multicolumn{2}{c}{\textbf{SMP}} \\ 
        \cmidrule(l){2-3} \cmidrule(l){4-5}
            &\textbf{Training}&\textbf{Inference}
            &\textbf{Training}& \textbf{Inference}\\         
    \midrule
        {FuseDNN, Ding \textit{et al.}'19}& 2 & 1 & 3 & 0.5\\
        {MMAtt, Xu \textit{et al.}'20} & 38 & 1 & 85 & 2\\
        VSCNN, Abousaleh \textit{et al.}'20 & 7 & 1& 23 & 2\\
        {HashPop, Liao \textit{et al.}'22} & 11& 1 &21  & 2\\
        {TweetGage, Arazzi\textit{et al.}'23} &  88 & 17 &188  & 62\\
        {MFTM, Hsu \textit{et al.}'23} & 6.5 &2.5  &7  & 3 \\
        ECSF, Mao \textit{et al.}'23 & 1.5  & 0.01 & 17 & 1 \\
        \textbf{NARRATOR} &\textbf{81} &\textbf{5} 
        &\textbf{175} &\textbf{14} \\
    \bottomrule    
    \end{tabular*}   
\end{table}
The training and inference time taken by different methods as listed in \autoref{table:res_time} is reported in seconds. As can be seen from \autoref{table:res_time}, NARRATOR's training time is notably higher than the state-of-the-art methods due to the complexity of handling multiple modalities and the substantial
number of trainable parameters. However, it is worth noting that its inference time is
significantly faster than a recent state-of-the-art method, TweetGage \cite{arazzi2023predicting}. This observation is atttibuted to the complexity of handling multiple modalities and the substantial number of trainable parameters in NARRATOR. Specifically, training on the TPIC dataset takes ~81 seconds per epoch and increases to ~175 seconds per epoch for the SMP dataset. In contrast, inference takes ~5 seconds and ~14 seconds per epoch for TPIC and SMP datasets, respectively. This suggests the model is capable of real-time or near-real-time predictions, a critical factor for many practical applications. NARRATOR's complexity is underscored by its memory footprint. With 34,903,017 trainable parameters, it occupies 133.14 MB of memory, highlighting its capacity to capture intricate patterns from multimodal data. Although NARRATOR incurs a higher computational cost compared to simpler baselines, its superior performance in predicting social media post popularity justifies the trade-off. The model's ability to leverage diverse modalities, including user demographics, sentiment analysis, and hashtag-guided attention, enables it to capture subtle nuances that contribute to post engagement. Future work could explore optimizations to enhance NARRATOR's computational efficiency without compromising its predictive accuracy, further broadening its applicability.
\section{Discussion}
This section delves into analysis of PCA on social features, comparison of novel hashtag-guided attention with parallel co-attention mechanism. Further, we discuss the limitations of NARRATOR and its broader significance, encompassing both theoretical and practical implications.
\subsection{Analysis of PCA on Social Features}
A simple technique such as PCA might seem out of place alongside the complex multimodal feature processing in our approach. However, feature extraction and dimensionality reduction are not contradictory but rather complementary steps. We first extract a rich set of features to capture the diverse aspects of the data, then use PCA to create a more manageable and efficient representation for the model. The motivation behind applying PCA was twofold:
\begin{enumerate}
\item Managing High Dimensionality: The extracted social features can be numerous, and some may introduce noise. PCA helps identify and retain the most important features while mitigating noise and reducing computational complexity.
\item Reducing Model Complexity: Applying PCA to social features significantly decreases the total number of trainable parameters from 197,268,457 to 34,903,017, and the memory usage from 752.52 MB to 133.14 MB. This reduction enhances the model's efficiency and scalability, crucial factors for real-world applications.
\end{enumerate}
While PCA is a relatively simple technique, our empirical results demonstrate its effectiveness in this context.
To further validate this, we conducted experiments comparing the performance of NARRATOR with and without PCA applied to social features. The results, presented below, show that while removing PCA leads to a slight decrease in performance and provides the benefits of reduced dimensionality and fewer trainable parameters.
\begin{table}[!ht] \footnotesize
\centering
\caption{Analysis of Dimensionality Reduction on PCA}
\label{table:res_pca_social}   
    \begin{tabular*}{0.9\linewidth}{@{\extracolsep{\fill}} lcccc }
    \toprule
        \textbf{Variant} & \multicolumn{2}{c}{\textbf{TPIC}} & \multicolumn{2}{c}{\textbf{SMP}} \\ 
        \cmidrule(l){2-3} \cmidrule(l){4-5}
            &\textbf{MSE}&\textbf{MAE}
            &\textbf{MSE}& \textbf{MAE}\\    
    \midrule
NARRATOR w/o PCA on Social & 1.495 & 0.913 & 2.616  & 1.267\\
NARRATOR with PCA on Social & 1.196 & 0.854 & 2.022 & 0.972\\
\midrule
   NARRATOR w/o Social & 2.166 & 1.093 & 2.897  & 1.267\\
   NARRATOR with Social  & 1.196 & 0.854 & 2.022 & 0.972\\
    \bottomrule  
    \end{tabular*}   
\end{table}
Furthermore, we conducted experiments by removing social features to investigate their importance. our ablation study results show a significant drop in performance when social features are discarded, as shown in Table~\ref{table:res_pca_social} underscores their valuable contribution to capturing the social context and dynamics that influence content popularity.
In conclusion, PCA strikes a suitable balance between dimensionality reduction, information preservation, and model complexity. While we acknowledge the potential of exploring alternative dimensionality reduction techniques in future work, our current approach effectively leverages social features while maintaining computational efficiency.
\subsection{Comparison with Parallel Co-attention}
In social media popularity prediction, understanding the interplay between various modalities is crucial, and while both hashtag-guided attention and parallel co-attention address this, they differ in how they incorporate contextual information.
\begin{figure}[!ht]
    \centering
    \subfloat[\centering Parallel Co-attention Mechanism]{{\includegraphics[width=6cm,height=12cm,keepaspectratio]{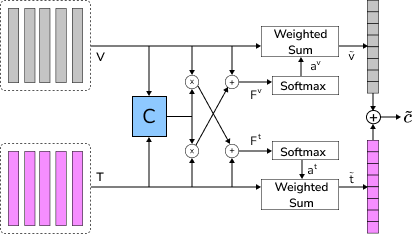} }}%
    \qquad
    \subfloat[\centering Hashtag-guided Attention Mechanism (Proposed)]{{\includegraphics[width=5cm]{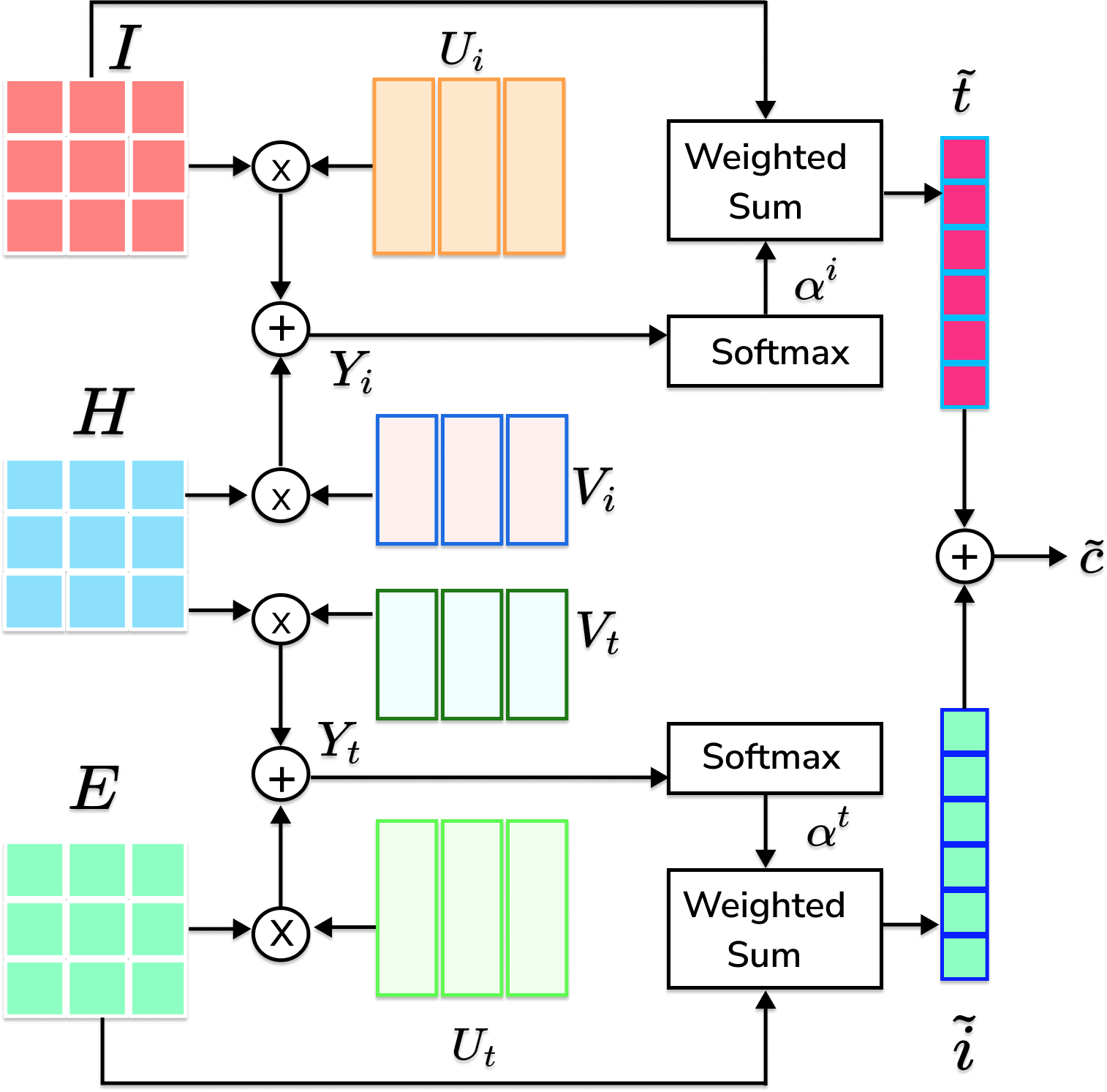} }}%
    \caption{Comparative Analysis of Attention Mechanisms}%
    \label{fig:example}%
\end{figure}
Our proposed hashtag-guided attention mechanism, illustrated in Figure \ref{fig:example}, draws inspiration from parallel co-attention but introduces a key innovation: the explicit utilization of hashtags as an additional source of context to compute attention weights. We recognize that the popularity of a post is not solely determined by its intrinsic content (text and images), but also by the broader social context in which it is shared, as conveyed through hashtags. By mathematically integrating hashtag embeddings into the attention computation, our mechanism dynamically weighs features from both textual and visual modalities based on the contextual cues provided by the hashtags. This approach fills a research gap by directly addressing the limitation of existing attention mechanisms, including parallel co-attention \cite{zhang2019hashtag}, which primarily focus on intrinsic content features and overlook the rich contextual cues offered by hashtags. By incorporating hashtag information, our mechanism guides attention and enhances feature selection, leading to improved prediction accuracy and interpretability. While the core concept of attention is not new, our specific implementation and integration of hashtag guidance into the attention mechanism represent a methodological innovation. We have carefully designed the algorithm to capture the complex interplay between hashtags, textual content, and visual content, resulting in a more effective and context-aware model. The key advantages of hashtag-guided attention are discussed below:
\begin{itemize}
\item Explicit Contextualization: Directly incorporates the semantic and contextual information encoded in hashtags, enhancing the model's understanding of the post's intended audience and relevance to trending topics.
\item Dynamic Adaptation: Allows the model to adapt its focus based on the specific hashtags associated with the post, making it more versatile for diverse content types and social media platforms.
\item Interpretability: Provides insights into how hashtags influence feature selection, offering greater transparency into the factors driving popularity predictions.
\end{itemize}
In contrast, parallel co-attention primarily focuses on capturing the interplay between textual and visual modalities within the post itself. It learns to align and weigh features from both modalities to generate a more holistic representation of the content. While valuable for understanding the intrinsic relationship between text and images, parallel co-attention neglects the broader social context provided by hashtags, which serve as crucial signals of the creator's intent, the target audience, and the current trends. As shown in Table \ref{table:att}, our hashtag-guided attention mechanism outperforms parallel co-attention, demonstrating an absolute improvement of 13.89\% and 56.35\% on TPIC and 39.95\% and 39.81\% on SMP in terms of MSE and MAE metrics, respectively. This highlights the effectiveness of incorporating hashtag information into the attention mechanism for improved popularity prediction. Both hashtag-guided attention and parallel co-attention contribute to a richer understanding of multimodal social media posts. However, by explicitly incorporating and mathematically integrating hashtag information into the attention computation, our hashtag-guided attention mechanism provides a more context-aware and adaptable approach, leading to more accurate and interpretable popularity predictions.
\subsection{Limitations and Future Work}
While NARRATOR exhibits several strengths, it is essential to acknowledge its limitations. The model's performance is inherently dependent on the quality and diversity of the training data. In scenarios where the training data is biased, incomplete, or fails to adequately represent the target population or specific social media platform, NARRATOR's predictions may exhibit biases. The accuracy of demographic information derived from faces in images may be compromised by factors such as image resolution, lighting conditions, image noise, and occlusions. Low-quality images or images with obscured faces may lead to inaccurate demographic predictions. Although NARRATOR demonstrates improved performance on real-world datasets, its applicability to other domains has yet to be explored. Cultural disparities, language variations, and platform-specific trends are all potential influences on the model's performance.
\subsection{Significance}
Accurately predicting online content popularity early on holds both theoretical and practical significance across various domains. To showcase the importance of popularity prediction and the capabilities of NARRATOR, we will delve into its theoretical implications and explore its potential real-world applications.
\subsubsection{Theoretical Implications}
NARRATOR's innovative approach to multimodal popularity prediction offers several key theoretical contributions, enhancing our understanding of content dynamics and user engagement on social media platforms.
\begin{enumerate}
\item Enhanced Understanding of Multimodal Content: NARRATOR's integration of visual demographics, analysing sentiments of hashtags, and hashtag-guided attention offers a better understanding of how different modalities (text, images, hashtags) interact to influence post popularity. This contributes to the theoretical understanding of multimodal content processing and its impact on user engagement.
\item Context-Aware Feature Selection: The hashtag-guided attention mechanism demonstrates the importance of incorporating contextual cues for effective feature selection in multimodal data. This highlights the limitations of traditional attention mechanisms that solely focus on intrinsic content features and opens new avenues for research into context-aware deep learning methods and techniques.
\item Role of Hashtag Sentiment in Popularity Prediction: NARRATOR's explicit incorporation of sentiment analysis from hashtags underscores their significance beyond mere metadata. By demonstrating the impact of hashtag sentiment on understanding audience perception and predicting post popularity, this research contributes to the theoretical understanding of how emotions expressed through hashtags shape online discourse and user engagement.
\end{enumerate}
\subsubsection{Practical Implications}
As discussed below, the practical implications of NARRATOR in multimodal popularity prediction are far-reaching, revolutionizing how individuals interact with social media platforms.
\begin{enumerate}
\item Improved Content Recommendation and Targeted Advertising: NARRATOR's ability to accurately predict popularity can enable social media platforms to recommend content that is more likely to resonate with individual users, leading to increased user engagement and satisfaction. Similarly, advertisers can benefit from more targeted ad placements based on predicted popularity and audience demographics.
\item Trend Detection and Analysis: NARRATOR's ability to analyze sentiment of hashtags, visual demographics and content-based features in conjunction with hashtag context can help researchers and marketers identify emerging trends, understand audience preferences, and track shifts in public opinion.
\item Enhanced User Experience: By facilitating the discovery of relevant and engaging content, NARRATOR can help mitigate information overload and enhance the overall user experience on social media platforms.
\item Combating Misinformation: The ability to predict the popularity of content early on can enable platform managers to proactively identify and address the spread of misinformation or harmful content.
\end{enumerate}
\section{Conclusion}
\label{sec:conclusions}
In this paper, we propose a novel paradigm for forecasting the popularity of social media posts by leveraging multimodal characteristics. Our approach leverages a multifaceted feature extraction process, capturing content-based information from both text and visuals, sentiment-oriented information from hashtags and text, user demographics, and social network data, along with topical and structural characteristics derived from hashtags. We propose a novel hashtag-guided attention mechanism that captures the influence of hashtags on both visual and textual content. This mechanism facilitates the model in learning the relative importance of different image regions and text segments based on their association with hashtags, leading to a more nuanced understanding of how hashtags shape user engagement. To demonstrate the efficacy of our proposed method, we undertake quantitative and qualitative comparisons in addition to ablation and statistical investigations. Our method achieves significant performance improvements compared to existing state-of-the-art approaches, as evaluated
on two real-world datasets. This finding suggests the potential effectiveness of our proposed method for predicting the popularity of multimodal posts.
\section*{Declarations}
\subsection*{Competing interests}
The authors declare that they have no known competing financial interests or personal relationships that could have appeared to influence the work reported in this paper.
\subsection*{Availability of data and materials}
Data will be made available on request.
\bibliography{bibliography}


\begin{thebibliography}{64}
\ifx \bisbn   \undefined \def \bisbn  #1{ISBN #1}\fi
\ifx \binits  \undefined \def \binits#1{#1}\fi
\ifx \bauthor  \undefined \def \bauthor#1{#1}\fi
\ifx \batitle  \undefined \def \batitle#1{#1}\fi
\ifx \bjtitle  \undefined \def \bjtitle#1{#1}\fi
\ifx \bvolume  \undefined \def \bvolume#1{\textbf{#1}}\fi
\ifx \byear  \undefined \def \byear#1{#1}\fi
\ifx \bissue  \undefined \def \bissue#1{#1}\fi
\ifx \bfpage  \undefined \def \bfpage#1{#1}\fi
\ifx \blpage  \undefined \def \blpage #1{#1}\fi
\ifx \burl  \undefined \def \burl#1{\textsf{#1}}\fi
\ifx \doiurl  \undefined \def \doiurl#1{\url{https://doi.org/#1}}\fi
\ifx \betal  \undefined \def \betal{\textit{et al.}}\fi
\ifx \binstitute  \undefined \def \binstitute#1{#1}\fi
\ifx \binstitutionaled  \undefined \def \binstitutionaled#1{#1}\fi
\ifx \bctitle  \undefined \def \bctitle#1{#1}\fi
\ifx \beditor  \undefined \def \beditor#1{#1}\fi
\ifx \bpublisher  \undefined \def \bpublisher#1{#1}\fi
\ifx \bbtitle  \undefined \def \bbtitle#1{#1}\fi
\ifx \bedition  \undefined \def \bedition#1{#1}\fi
\ifx \bseriesno  \undefined \def \bseriesno#1{#1}\fi
\ifx \blocation  \undefined \def \blocation#1{#1}\fi
\ifx \bsertitle  \undefined \def \bsertitle#1{#1}\fi
\ifx \bsnm \undefined \def \bsnm#1{#1}\fi
\ifx \bsuffix \undefined \def \bsuffix#1{#1}\fi
\ifx \bparticle \undefined \def \bparticle#1{#1}\fi
\ifx \barticle \undefined \def \barticle#1{#1}\fi
\bibcommenthead
\ifx \bconfdate \undefined \def \bconfdate #1{#1}\fi
\ifx \botherref \undefined \def \botherref #1{#1}\fi
\ifx \url \undefined \def \url#1{\textsf{#1}}\fi
\ifx \bchapter \undefined \def \bchapter#1{#1}\fi
\ifx \bbook \undefined \def \bbook#1{#1}\fi
\ifx \bcomment \undefined \def \bcomment#1{#1}\fi
\ifx \oauthor \undefined \def \oauthor#1{#1}\fi
\ifx \citeauthoryear \undefined \def \citeauthoryear#1{#1}\fi
\ifx \endbibitem  \undefined \def \endbibitem {}\fi
\ifx \bconflocation  \undefined \def \bconflocation#1{#1}\fi
\ifx \arxivurl  \undefined \def \arxivurl#1{\textsf{#1}}\fi
\csname PreBibitemsHook\endcsname

\bibitem[\protect\citeauthoryear{Shen et~al.}{2020}]{shen2020examining}
\begin{barticle}
\bauthor{\bsnm{Shen}, \binits{F.}},
\bauthor{\bsnm{Xia}, \binits{C.}},
\bauthor{\bsnm{Skoric}, \binits{M.}}:
\batitle{Examining the roles of social media and alternative media in social movement participation: A study of hong kong’s umbrella movement}.
\bjtitle{Telematics and Informatics}
\bvolume{47},
\bfpage{101303}
(\byear{2020})
\doiurl{10.1016/j.tele.2019.101303}
\end{barticle}
\endbibitem

\bibitem[\protect\citeauthoryear{Anderson and Brook}{2021}]{Anderson2021}
\begin{botherref}
\oauthor{\bsnm{Anderson}, \binits{M.}},
\oauthor{\bsnm{Brook}, \binits{A.}}:
Social Media Use in 2021 | Pew Research Center
(2021)
\end{botherref}
\endbibitem

\bibitem[\protect\citeauthoryear{Cao et~al.}{2020}]{cao2020popularity}
\begin{bchapter}
\bauthor{\bsnm{Cao}, \binits{Q.}},
\bauthor{\bsnm{Shen}, \binits{H.}},
\bauthor{\bsnm{Gao}, \binits{J.}},
\bauthor{\bsnm{Wei}, \binits{B.}},
\bauthor{\bsnm{Cheng}, \binits{X.}}:
\bctitle{Popularity prediction on social platforms with coupled graph neural networks}.
In: \bbtitle{Proceedings of the 13th International Conference on Web Search and Data Mining},
pp. \bfpage{70}--\blpage{78}
(\byear{2020}).
\doiurl{10.1145/3336191.3371834}
\end{bchapter}
\endbibitem

\bibitem[\protect\citeauthoryear{Wu et~al.}{2016}]{wu2016unfolding}
\begin{bchapter}
\bauthor{\bsnm{Wu}, \binits{B.}},
\bauthor{\bsnm{Mei}, \binits{T.}},
\bauthor{\bsnm{Cheng}, \binits{W.-H.}},
\bauthor{\bsnm{Zhang}, \binits{Y.}}:
\bctitle{Unfolding temporal dynamics: Predicting social media popularity using multi-scale temporal decomposition}.
In: \bbtitle{Proceedings of the AAAI Conference on Artificial Intelligence},
vol. \bseriesno{30}
(\byear{2016}).
\doiurl{10.1609/aaai.v30i1.9970}
\end{bchapter}
\endbibitem

\bibitem[\protect\citeauthoryear{Gon{\c{c}}alves et~al.}{2010}]{gonccalves2010popularity}
\begin{barticle}
\bauthor{\bsnm{Gon{\c{c}}alves}, \binits{M.A.}},
\bauthor{\bsnm{Almeida}, \binits{J.M.}},
\bauthor{\bsnm{Santos}, \binits{L.G.}},
\bauthor{\bsnm{Laender}, \binits{A.H.}},
\bauthor{\bsnm{Almeida}, \binits{V.}}:
\batitle{On popularity in the blogosphere}.
\bjtitle{IEEE Internet Computing}
\bvolume{14}(\bissue{3}),
\bfpage{42}--\blpage{49}
(\byear{2010})
\doiurl{10.1109/MIC.2010.73}
\end{barticle}
\endbibitem

\bibitem[\protect\citeauthoryear{Majid et~al.}{2013}]{majid2013context}
\begin{barticle}
\bauthor{\bsnm{Majid}, \binits{A.}},
\bauthor{\bsnm{Chen}, \binits{L.}},
\bauthor{\bsnm{Chen}, \binits{G.}},
\bauthor{\bsnm{Mirza}, \binits{H.T.}},
\bauthor{\bsnm{Hussain}, \binits{I.}},
\bauthor{\bsnm{Woodward}, \binits{J.}}:
\batitle{A context-aware personalized travel recommendation system based on geotagged social media data mining}.
\bjtitle{International Journal of Geographical Information Science}
\bvolume{27}(\bissue{4}),
\bfpage{662}--\blpage{684}
(\byear{2013})
\doiurl{10.1080/13658816.2012.696649}
\end{barticle}
\endbibitem

\bibitem[\protect\citeauthoryear{Li et~al.}{2015}]{li2015click}
\begin{bchapter}
\bauthor{\bsnm{Li}, \binits{C.}},
\bauthor{\bsnm{Lu}, \binits{Y.}},
\bauthor{\bsnm{Mei}, \binits{Q.}},
\bauthor{\bsnm{Wang}, \binits{D.}},
\bauthor{\bsnm{Pandey}, \binits{S.}}:
\bctitle{Click-through prediction for advertising in twitter timeline}.
In: \bbtitle{Proceedings of the 21th ACM SIGKDD International Conference on Knowledge Discovery and Data Mining},
pp. \bfpage{1959}--\blpage{1968}
(\byear{2015}).
\doiurl{10.1145/2783258.2788582}
\end{bchapter}
\endbibitem

\bibitem[\protect\citeauthoryear{Aven et~al.}{2014}]{aven2014using}
\begin{botherref}
\oauthor{\bsnm{Aven}, \binits{B.L.}},
\oauthor{\bsnm{Burgess}, \binits{D.A.}},
\oauthor{\bsnm{Haynes}, \binits{J.F.}},
\oauthor{\bsnm{Merino}, \binits{J.R.}},
\oauthor{\bsnm{Moore}, \binits{P.C.}}:
Using product and social network data to improve online advertising.
Google Patents.
US Patent 8,843,406
(2014)
\end{botherref}
\endbibitem

\bibitem[\protect\citeauthoryear{Roy et~al.}{2013}]{roy2013towards}
\begin{barticle}
\bauthor{\bsnm{Roy}, \binits{S.D.}},
\bauthor{\bsnm{Mei}, \binits{T.}},
\bauthor{\bsnm{Zeng}, \binits{W.}},
\bauthor{\bsnm{Li}, \binits{S.}}:
\batitle{Towards cross-domain learning for social video popularity prediction}.
\bjtitle{IEEE Transactions on multimedia}
\bvolume{15}(\bissue{6}),
\bfpage{1255}--\blpage{1267}
(\byear{2013})
\doiurl{10.1109/TMM.2013.2265079}
\end{barticle}
\endbibitem

\bibitem[\protect\citeauthoryear{Gan et~al.}{2016}]{gan2016webly}
\begin{bchapter}
\bauthor{\bsnm{Gan}, \binits{C.}},
\bauthor{\bsnm{Sun}, \binits{C.}},
\bauthor{\bsnm{Duan}, \binits{L.}},
\bauthor{\bsnm{Gong}, \binits{B.}}:
\bctitle{Webly-supervised video recognition by mutually voting for relevant web images and web video frames}.
In: \bbtitle{Computer Vision--ECCV 2016: 14th European Conference, Amsterdam, The Netherlands, October 11-14, 2016, Proceedings, Part III 14},
pp. \bfpage{849}--\blpage{866}
(\byear{2016}).
\doiurl{10.1007/978-3-319-46487-9_52} .
\bcomment{Springer}
\end{bchapter}
\endbibitem

\bibitem[\protect\citeauthoryear{Kim et~al.}{2015}]{kim2015demand}
\begin{barticle}
\bauthor{\bsnm{Kim}, \binits{W.}},
\bauthor{\bsnm{Won}, \binits{J.H.}},
\bauthor{\bsnm{Park}, \binits{S.}},
\bauthor{\bsnm{Kang}, \binits{J.}}:
\batitle{Demand forecasting models for medicines through wireless sensor networks data and topic trend analysis}.
\bjtitle{International Journal of Distributed Sensor Networks}
\bvolume{11}(\bissue{9}),
\bfpage{907169}
(\byear{2015})
\doiurl{10.1155/2015/907169}
\end{barticle}
\endbibitem

\bibitem[\protect\citeauthoryear{Wang et~al.}{2021}]{wang2021deep}
\begin{bchapter}
\bauthor{\bsnm{Wang}, \binits{J.}},
\bauthor{\bsnm{Xu}, \binits{B.}},
\bauthor{\bsnm{Zu}, \binits{Y.}}:
\bctitle{Deep learning for aspect-based sentiment analysis}.
In: \bbtitle{2021 International Conference on Machine Learning and Intelligent Systems Engineering (MLISE)},
pp. \bfpage{267}--\blpage{271}
(\byear{2021}).
\doiurl{10.1109/MLISE54096.2021.00056} .
\bcomment{IEEE}
\end{bchapter}
\endbibitem

\bibitem[\protect\citeauthoryear{Saura}{2021}]{saura2021using}
\begin{barticle}
\bauthor{\bsnm{Saura}, \binits{J.R.}}:
\batitle{Using data sciences in digital marketing: Framework, methods, and performance metrics}.
\bjtitle{Journal of Innovation \& Knowledge}
\bvolume{6}(\bissue{2}),
\bfpage{92}--\blpage{102}
(\byear{2021})
\doiurl{10.1016/j.jik.2020.08.001}
\end{barticle}
\endbibitem

\bibitem[\protect\citeauthoryear{Saura et~al.}{2021}]{saura2021user}
\begin{barticle}
\bauthor{\bsnm{Saura}, \binits{J.R.}},
\bauthor{\bsnm{Ribeiro-Soriano}, \binits{D.}},
\bauthor{\bsnm{Palacios-Marqu{\'e}s}, \binits{D.}}:
\batitle{From user-generated data to data-driven innovation: A research agenda to understand user privacy in digital markets}.
\bjtitle{International Journal of Information Management}
\bvolume{60},
\bfpage{102331}
(\byear{2021})
\doiurl{10.1016/j.ijinfomgt.2021.102331}
\end{barticle}
\endbibitem

\bibitem[\protect\citeauthoryear{Ribeiro-Navarrete et~al.}{2021}]{ribeiro2021towards}
\begin{barticle}
\bauthor{\bsnm{Ribeiro-Navarrete}, \binits{S.}},
\bauthor{\bsnm{Saura}, \binits{J.R.}},
\bauthor{\bsnm{Palacios-Marqu{\'e}s}, \binits{D.}}:
\batitle{Towards a new era of mass data collection: Assessing pandemic surveillance technologies to preserve user privacy}.
\bjtitle{Technological Forecasting and Social Change}
\bvolume{167},
\bfpage{120681}
(\byear{2021})
\doiurl{10.1016/j.techfore.2021.120681}
\end{barticle}
\endbibitem

\bibitem[\protect\citeauthoryear{Xu et~al.}{2020}]{xu2020multimodal}
\begin{bchapter}
\bauthor{\bsnm{Xu}, \binits{K.}},
\bauthor{\bsnm{Lin}, \binits{Z.}},
\bauthor{\bsnm{Zhao}, \binits{J.}},
\bauthor{\bsnm{Shi}, \binits{P.}},
\bauthor{\bsnm{Deng}, \binits{W.}},
\bauthor{\bsnm{Wang}, \binits{H.}}:
\bctitle{Multimodal deep learning for social media popularity prediction with attention mechanism}.
In: \bbtitle{MM 2020 - Proceedings of the 28th ACM International Conference on Multimedia}
(\byear{2020}).
\doiurl{10.1145/3394171.3416274}
\end{bchapter}
\endbibitem

\bibitem[\protect\citeauthoryear{Lin et~al.}{2022}]{Lin2022}
\begin{botherref}
\oauthor{\bsnm{Lin}, \binits{H.H.}},
\oauthor{\bsnm{Lin}, \binits{J.D.}},
\oauthor{\bsnm{Ople}, \binits{J.J.M.}},
\oauthor{\bsnm{Chen}, \binits{J.C.}},
\oauthor{\bsnm{Hua}, \binits{K.L.}}:
Social media popularity prediction based on multi-modal self-attention mechanisms.
IEEE Access
\textbf{10}
(2022)
\doiurl{10.1109/ACCESS.2021.3136552}
\end{botherref}
\endbibitem

\bibitem[\protect\citeauthoryear{Nguyen et~al.}{2019}]{nguyen2019attention}
\begin{bchapter}
\bauthor{\bsnm{Nguyen}, \binits{M.-T.}},
\bauthor{\bsnm{Le}, \binits{D.H.}},
\bauthor{\bsnm{Nakajima}, \binits{T.}},
\bauthor{\bsnm{Yoshimi}, \binits{M.}},
\bauthor{\bsnm{Thoai}, \binits{N.}}:
\bctitle{Attention-based neural network: A novel approach for predicting the popularity of online content}.
In: \bbtitle{2019 IEEE 21st International Conference on High Performance Computing and Communications; IEEE 17th International Conference on Smart City; IEEE 5th International Conference on Data Science and Systems (HPCC/SmartCity/DSS)},
pp. \bfpage{329}--\blpage{336}
(\byear{2019}).
\doiurl{10.1109/HPCC/SmartCity/DSS.2019.00058} .
\bcomment{IEEE}
\end{bchapter}
\endbibitem

\bibitem[\protect\citeauthoryear{Liao et~al.}{2019}]{liao2019popularity}
\begin{bchapter}
\bauthor{\bsnm{Liao}, \binits{D.}},
\bauthor{\bsnm{Xu}, \binits{J.}},
\bauthor{\bsnm{Li}, \binits{G.}},
\bauthor{\bsnm{Huang}, \binits{W.}},
\bauthor{\bsnm{Liu}, \binits{W.}},
\bauthor{\bsnm{Li}, \binits{J.}}:
\bctitle{Popularity prediction on online articles with deep fusion of temporal process and content features}.
In: \bbtitle{Proceedings of the AAAI Conference on Artificial Intelligence},
vol. \bseriesno{33},
pp. \bfpage{200}--\blpage{207}
(\byear{2019}).
\doiurl{10.1609/aaai.v33i01.3301200}
\end{bchapter}
\endbibitem

\bibitem[\protect\citeauthoryear{Chen et~al.}{2019}]{chen2019social}
\begin{bchapter}
\bauthor{\bsnm{Chen}, \binits{J.}},
\bauthor{\bsnm{Liang}, \binits{D.}},
\bauthor{\bsnm{Zhu}, \binits{Z.}},
\bauthor{\bsnm{Zhou}, \binits{X.}},
\bauthor{\bsnm{Ye}, \binits{Z.}},
\bauthor{\bsnm{Mo}, \binits{X.}}:
\bctitle{Social media popularity prediction based on visual-textual features with xgboost}.
In: \bbtitle{Proceedings of the 27th ACM International Conference on Multimedia},
pp. \bfpage{2692}--\blpage{2696}
(\byear{2019}).
\doiurl{10.1145/3343031.335607}
\end{bchapter}
\endbibitem

\bibitem[\protect\citeauthoryear{Zhang et~al.}{2018a}]{zhang2018become}
\begin{bchapter}
\bauthor{\bsnm{Zhang}, \binits{Z.}},
\bauthor{\bsnm{Chen}, \binits{T.}},
\bauthor{\bsnm{Zhou}, \binits{Z.}},
\bauthor{\bsnm{Li}, \binits{J.}},
\bauthor{\bsnm{Luo}, \binits{J.}}:
\bctitle{How to become instagram famous: Post popularity prediction with dual-attention}.
In: \bbtitle{2018 IEEE International Conference on Big Data (big Data)},
pp. \bfpage{2383}--\blpage{2392}
(\byear{2018}).
\doiurl{10.1109/BigData.2018.8622461} .
\bcomment{IEEE}
\end{bchapter}
\endbibitem

\bibitem[\protect\citeauthoryear{Zhang et~al.}{2018b}]{zhang2018user}
\begin{bchapter}
\bauthor{\bsnm{Zhang}, \binits{W.}},
\bauthor{\bsnm{Wang}, \binits{W.}},
\bauthor{\bsnm{Wang}, \binits{J.}},
\bauthor{\bsnm{Zha}, \binits{H.}}:
\bctitle{User-guided hierarchical attention network for multi-modal social image popularity prediction}.
In: \bbtitle{The Web Conference 2018 - Proceedings of the World Wide Web Conference, WWW 2018}
(\byear{2018}).
\doiurl{10.1145/3178876.3186026}
\end{bchapter}
\endbibitem

\bibitem[\protect\citeauthoryear{Wang et~al.}{2023}]{wang2023social}
\begin{barticle}
\bauthor{\bsnm{Wang}, \binits{J.}},
\bauthor{\bsnm{Yang}, \binits{S.}},
\bauthor{\bsnm{Zhao}, \binits{H.}},
\bauthor{\bsnm{Yang}, \binits{Y.}}:
\batitle{Social media popularity prediction with multimodal hierarchical fusion model}.
\bjtitle{Computer Speech \& Language}
\bvolume{80},
\bfpage{101490}
(\byear{2023})
\doiurl{10.1016/j.csl.2023.101490}
\end{barticle}
\endbibitem

\bibitem[\protect\citeauthoryear{Caleffi}{2015}]{Caleffi2015}
\begin{botherref}
\oauthor{\bsnm{Caleffi}, \binits{P.-M.}}:
The `hashtag': A new word or a new rule?
SKASE Journal of Theoretical Linguistics
\textbf{12}
(2015)
\doiurl{10.24093/awej/call6.6}
\end{botherref}
\endbibitem

\bibitem[\protect\citeauthoryear{Zhang et~al.}{2019}]{zhang2019hashtag}
\begin{bchapter}
\bauthor{\bsnm{Zhang}, \binits{S.}},
\bauthor{\bsnm{Yao}, \binits{Y.}},
\bauthor{\bsnm{Xu}, \binits{F.}},
\bauthor{\bsnm{Tong}, \binits{H.}},
\bauthor{\bsnm{Yan}, \binits{X.}},
\bauthor{\bsnm{Lu}, \binits{J.}}:
\bctitle{Hashtag recommendation for photo sharing services}.
In: \bbtitle{Proceedings of the AAAI Conference on Artificial Intelligence}
(\byear{2019}).
\doiurl{10.1609/aaai.v33i01.33015805}
\end{bchapter}
\endbibitem

\bibitem[\protect\citeauthoryear{Abousaleh et~al.}{2021}]{abousaleh2020multimodal}
\begin{botherref}
\oauthor{\bsnm{Abousaleh}, \binits{F.S.}},
\oauthor{\bsnm{Cheng}, \binits{W.H.}},
\oauthor{\bsnm{Yu}, \binits{N.H.}},
\oauthor{\bsnm{Tsao}, \binits{Y.}}:
Multimodal deep learning framework for image popularity prediction on social media.
IEEE Transactions on Cognitive and Developmental Systems
\textbf{13}
(2021)
\doiurl{10.1109/TCDS.2020.3036690}
\end{botherref}
\endbibitem

\bibitem[\protect\citeauthoryear{Bakhshi et~al.}{2014}]{bakhshi2014faces}
\begin{bchapter}
\bauthor{\bsnm{Bakhshi}, \binits{S.}},
\bauthor{\bsnm{Shamma}, \binits{D.A.}},
\bauthor{\bsnm{Gilbert}, \binits{E.}}:
\bctitle{Faces engage us: Photos with faces attract more likes and comments on instagram}.
In: \bbtitle{Proceedings of the SIGCHI Conference on Human Factors in Computing Systems}
(\byear{2014}).
\doiurl{10.1145/2556288.2557403}
\end{bchapter}
\endbibitem

\bibitem[\protect\citeauthoryear{Gelli et~al.}{2015}]{gelli2015image}
\begin{bchapter}
\bauthor{\bsnm{Gelli}, \binits{F.}},
\bauthor{\bsnm{Uricchio}, \binits{T.}},
\bauthor{\bsnm{Bertini}, \binits{M.}},
\bauthor{\bsnm{Bimbo}, \binits{A.D.}},
\bauthor{\bsnm{Chang}, \binits{S.F.}}:
\bctitle{Image popularity prediction in social media using sentiment and context features}.
In: \bbtitle{Proceedings of the 23rd ACM International Conference on Multimedia}
(\byear{2015}).
\doiurl{10.1145/2733373.2806361}
\end{bchapter}
\endbibitem

\bibitem[\protect\citeauthoryear{Li et~al.}{2019}]{Li2019}
\begin{bchapter}
\bauthor{\bsnm{Li}, \binits{J.}},
\bauthor{\bsnm{Gao}, \binits{Y.}},
\bauthor{\bsnm{Gao}, \binits{X.}},
\bauthor{\bsnm{Shi}, \binits{Y.}},
\bauthor{\bsnm{Chen}, \binits{G.}}:
\bctitle{Senti2pop: Sentiment-aware topic popularity prediction on social media}.
In: \bbtitle{Proceedings - IEEE International Conference on Data Mining, ICDM},
vol. \bseriesno{2019-November}
(\byear{2019}).
\doiurl{10.1109/ICDM.2019.00143}
\end{bchapter}
\endbibitem

\bibitem[\protect\citeauthoryear{Mannepalli et~al.}{2023}]{mannepalli2023popularity}
\begin{barticle}
\bauthor{\bsnm{Mannepalli}, \binits{K.}},
\bauthor{\bsnm{Singh}, \binits{S.P.}},
\bauthor{\bsnm{Kolli}, \binits{C.S.}},
\bauthor{\bsnm{Raj}, \binits{S.}},
\bauthor{\bsnm{Bojja}, \binits{G.R.}},
\bauthor{\bsnm{Rajakumar}, \binits{B.}},
\bauthor{\bsnm{Binu}, \binits{D.}}:
\batitle{Popularity prediction model with context, time and user sentiment information: An optimization assisted deep learning technique}.
\bjtitle{International Journal of Uncertainty, Fuzziness and Knowledge-Based Systems}
\bvolume{31}(\bissue{02}),
\bfpage{283}--\blpage{302}
(\byear{2023})
\doiurl{10.1142/S0218488523500150}
\end{barticle}
\endbibitem

\bibitem[\protect\citeauthoryear{Yang et~al.}{2020}]{yang2020sentiment}
\begin{barticle}
\bauthor{\bsnm{Yang}, \binits{C.}},
\bauthor{\bsnm{Wang}, \binits{X.}},
\bauthor{\bsnm{Jiang}, \binits{B.}}:
\batitle{Sentiment enhanced multi-modal hashtag recommendation for micro-videos}.
\bjtitle{IEEE Access}
\bvolume{8},
\bfpage{78252}--\blpage{78264}
(\byear{2020})
\doiurl{10.1109/ACCESS.2020.2989473}
\end{barticle}
\endbibitem

\bibitem[\protect\citeauthoryear{Liao}{2022}]{liao-2022-leveraging}
\begin{bchapter}
\bauthor{\bsnm{Liao}, \binits{Y.Y.}}:
\bctitle{Leveraging hashtag networks for multimodal popularity prediction of {I}nstagram posts}.
In: \beditor{\bsnm{Calzolari}, \binits{N.}},
\beditor{\bsnm{B{\'e}chet}, \binits{F.}},
\beditor{\bsnm{Blache}, \binits{P.}},
\beditor{\bsnm{Choukri}, \binits{K.}},
\beditor{\bsnm{Cieri}, \binits{C.}},
\beditor{\bsnm{Declerck}, \binits{T.}},
\beditor{\bsnm{Goggi}, \binits{S.}},
\beditor{\bsnm{Isahara}, \binits{H.}},
\beditor{\bsnm{Maegaard}, \binits{B.}},
\beditor{\bsnm{Mariani}, \binits{J.}},
\beditor{\bsnm{Mazo}, \binits{H.}},
\beditor{\bsnm{Odijk}, \binits{J.}},
\beditor{\bsnm{Piperidis}, \binits{S.}} (eds.)
\bbtitle{Proceedings of the Thirteenth Language Resources and Evaluation Conference},
pp. \bfpage{7191}--\blpage{7198}.
\bpublisher{European Language Resources Association},
\blocation{Marseille, France}
(\byear{2022}).
\burl{https://aclanthology.org/2022.lrec-1.779}
\end{bchapter}
\endbibitem

\bibitem[\protect\citeauthoryear{Arazzi et~al.}{2023}]{arazzi2023predicting}
\begin{bchapter}
\bauthor{\bsnm{Arazzi}, \binits{M.}},
\bauthor{\bsnm{Cotogni}, \binits{M.}},
\bauthor{\bsnm{Nocera}, \binits{A.}},
\bauthor{\bsnm{Virgili}, \binits{L.}}:
\bctitle{Predicting tweet engagement with graph neural networks}.
In: \bbtitle{Proceedings of the 2023 ACM International Conference on Multimedia Retrieval},
pp. \bfpage{172}--\blpage{180}
(\byear{2023}).
\doiurl{10.1145/3591106.3592294}
\end{bchapter}
\endbibitem

\bibitem[\protect\citeauthoryear{Purba et~al.}{2021}]{Purba2021}
\begin{botherref}
\oauthor{\bsnm{Purba}, \binits{K.R.}},
\oauthor{\bsnm{Asirvatham}, \binits{D.}},
\oauthor{\bsnm{Murugesan}, \binits{R.K.}}:
Instagram post popularity trend analysis and prediction using hashtag, image assessment, and user history features.
International Arab Journal of Information Technology
\textbf{18}
(2021)
\doiurl{10.34028/iajit/18/1/10}
\end{botherref}
\endbibitem

\bibitem[\protect\citeauthoryear{Kumar et~al.}{2017}]{Kumar2017}
\begin{bchapter}
\bauthor{\bsnm{Kumar}, \binits{N.}},
\bauthor{\bsnm{Yadandla}, \binits{A.}},
\bauthor{\bsnm{Suryamukhi}, \binits{K.}},
\bauthor{\bsnm{Ranabothu}, \binits{N.}},
\bauthor{\bsnm{Boya}, \binits{S.}},
\bauthor{\bsnm{Singh}, \binits{M.}}:
\bctitle{Arousal prediction of news articles in social media},
vol. \bseriesno{10682 LNAI}
(\byear{2017}).
\doiurl{10.1007/978-3-319-71928-3_30}
\end{bchapter}
\endbibitem

\bibitem[\protect\citeauthoryear{Lin et~al.}{2019}]{lin2019layer}
\begin{botherref}
\oauthor{\bsnm{Lin}, \binits{Z.}},
\oauthor{\bsnm{Huang}, \binits{F.}},
\oauthor{\bsnm{Li}, \binits{Y.}},
\oauthor{\bsnm{Yang}, \binits{Z.}},
\oauthor{\bsnm{Liu}, \binits{W.}}:
A layer-wise deep stacking model for social image popularity prediction.
World Wide Web
\textbf{22}
(2019)
\doiurl{10.1007/s11280-018-0590-1}
\end{botherref}
\endbibitem

\bibitem[\protect\citeauthoryear{Cao et~al.}{2020}]{cao2020ppsp}
\begin{bchapter}
\bauthor{\bsnm{Cao}, \binits{Q.}},
\bauthor{\bsnm{Shen}, \binits{H.}},
\bauthor{\bsnm{Gao}, \binits{J.}},
\bauthor{\bsnm{Wei}, \binits{B.}},
\bauthor{\bsnm{Cheng}, \binits{X.}}:
\bctitle{Popularity prediction on social platforms with coupled graph neural networks}.
(\byear{2020}).
\doiurl{10.1145/3336191.3371834}
\end{bchapter}
\endbibitem

\bibitem[\protect\citeauthoryear{Mannepalli et~al.}{2023}]{mannepalli2023ppm}
\begin{botherref}
\oauthor{\bsnm{Mannepalli}, \binits{K.}},
\oauthor{\bsnm{Singh}, \binits{S.P.}},
\oauthor{\bsnm{Kolli}, \binits{C.S.}},
\oauthor{\bsnm{Raj}, \binits{S.}},
\oauthor{\bsnm{Bojja}, \binits{G.R.}},
\oauthor{\bsnm{Rajakumar}, \binits{B.R.}},
\oauthor{\bsnm{Binu}, \binits{D.}}:
Popularity prediction model with context, time and user sentiment information: An optimization assisted deep learning technique.
International Journal of Uncertainty, Fuzziness and Knowledge-Based Systems
\textbf{31}
(2023)
\doiurl{10.1142/S0218488523500150}
\end{botherref}
\endbibitem

\bibitem[\protect\citeauthoryear{Tan et~al.}{2022}]{tan2022emv}
\begin{bchapter}
\bauthor{\bsnm{Tan}, \binits{Y.}},
\bauthor{\bsnm{Liu}, \binits{F.}},
\bauthor{\bsnm{Li}, \binits{B.}},
\bauthor{\bsnm{Zhang}, \binits{Z.}},
\bauthor{\bsnm{Zhang}, \binits{B.}}:
\bctitle{An efficient multi-view multimodal data processing framework for social media popularity prediction}.
In: \bbtitle{MM 2022 - Proceedings of the 30th ACM International Conference on Multimedia},
pp. \bfpage{7200}--\blpage{7204}
(\byear{2022}).
\doiurl{10.1145/3503161.3551607}
\end{bchapter}
\endbibitem

\bibitem[\protect\citeauthoryear{Zappavigna}{2015}]{zappavigna2015searchable}
\begin{barticle}
\bauthor{\bsnm{Zappavigna}, \binits{M.}}:
\batitle{Searchable talk: The linguistic functions of hashtags}.
\bjtitle{Social Semiotics}
\bvolume{25}(\bissue{3}),
\bfpage{274}--\blpage{291}
(\byear{2015})
\doiurl{10.1080/10350330.2014.996948}
\end{barticle}
\endbibitem

\bibitem[\protect\citeauthoryear{Liu et~al.}{2018}]{liu2018hashtag2vec}
\begin{bchapter}
\bauthor{\bsnm{Liu}, \binits{J.}},
\bauthor{\bsnm{He}, \binits{Z.}},
\bauthor{\bsnm{Huang}, \binits{Y.}}:
\bctitle{Hashtag2vec: Learning hashtag representation with relational hierarchical embedding model}.
In: \bbtitle{IJCAI},
vol. \bseriesno{2018-July}
(\byear{2018}).
\doiurl{10.24963/ijcai.2018/480}
\end{bchapter}
\endbibitem

\bibitem[\protect\citeauthoryear{Chakrabarti et~al.}{2023}]{chakrabarti2023hr}
\begin{botherref}
\oauthor{\bsnm{Chakrabarti}, \binits{P.}},
\oauthor{\bsnm{Malvi}, \binits{E.}},
\oauthor{\bsnm{Bansal}, \binits{S.}},
\oauthor{\bsnm{Kumar}, \binits{N.}}:
Hashtag recommendation for enhancing the popularity of social media posts.
Social Network Analysis and Mining
\textbf{13}
(2023)
\doiurl{10.1007/s13278-023-01024-9}
\end{botherref}
\endbibitem

\bibitem[\protect\citeauthoryear{Bansal et~al.}{2023}]{Bansal2023}
\begin{botherref}
\oauthor{\bsnm{Bansal}, \binits{S.}},
\oauthor{\bsnm{Gowda}, \binits{K.}},
\oauthor{\bsnm{Kumar}, \binits{N.}}:
A hybrid deep neural network for multimodal personalized hashtag recommendation.
IEEE Transactions on Computational Social Systems
\textbf{10}
(2023)
\doiurl{10.1109/TCSS.2022.3184307}
\end{botherref}
\endbibitem

\bibitem[\protect\citeauthoryear{Wang et~al.}{2023}]{wang2023sm}
\begin{botherref}
\oauthor{\bsnm{Wang}, \binits{J.}},
\oauthor{\bsnm{Yang}, \binits{S.}},
\oauthor{\bsnm{Zhao}, \binits{H.}},
\oauthor{\bsnm{Yang}, \binits{Y.}}:
Social media popularity prediction with multimodal hierarchical fusion model.
Computer Speech and Language
\textbf{80}
(2023)
\doiurl{10.1016/j.csl.2023.101490}
\end{botherref}
\endbibitem

\bibitem[\protect\citeauthoryear{Devlin et~al.}{2019}]{devlin-etal-2019-bert}
\begin{bchapter}
\bauthor{\bsnm{Devlin}, \binits{J.}},
\bauthor{\bsnm{Chang}, \binits{M.-W.}},
\bauthor{\bsnm{Lee}, \binits{K.}},
\bauthor{\bsnm{Toutanova}, \binits{K.}}:
\bctitle{{BERT}: Pre-training of deep bidirectional transformers for language understanding}.
In: \beditor{\bsnm{Burstein}, \binits{J.}},
\beditor{\bsnm{Doran}, \binits{C.}},
\beditor{\bsnm{Solorio}, \binits{T.}} (eds.)
\bbtitle{Proceedings of the 2019 Conference of the North {A}merican Chapter of the Association for Computational Linguistics: Human Language Technologies, Volume 1 (Long and Short Papers)},
pp. \bfpage{4171}--\blpage{4186}.
\bpublisher{Association for Computational Linguistics},
\blocation{Minneapolis, Minnesota}
(\byear{2019}).
\doiurl{10.18653/v1/N19-1423} .
\burl{https://aclanthology.org/N19-1423}
\end{bchapter}
\endbibitem

\bibitem[\protect\citeauthoryear{Mikolov et~al.}{2013}]{mikolov2013distributed}
\begin{botherref}
\oauthor{\bsnm{Mikolov}, \binits{T.}},
\oauthor{\bsnm{Sutskever}, \binits{I.}},
\oauthor{\bsnm{Chen}, \binits{K.}},
\oauthor{\bsnm{Corrado}, \binits{G.}},
\oauthor{\bsnm{Dean}, \binits{J.}}:
Distributed representations of words and phrases and their compositionality.
Advances in Neural Information Processing Systems
(2013)
\end{botherref}
\endbibitem

\bibitem[\protect\citeauthoryear{Simonyan and Zisserman}{2015}]{simonyan2014very}
\begin{bchapter}
\bauthor{\bsnm{Simonyan}, \binits{K.}},
\bauthor{\bsnm{Zisserman}, \binits{A.}}:
\bctitle{Very deep convolutional networks for large-scale image recognition}.
In: \bbtitle{3rd International Conference on Learning Representations, ICLR 2015 - Conference Track Proceedings}
(\byear{2015})
\end{bchapter}
\endbibitem

\bibitem[\protect\citeauthoryear{Deng et~al.}{2010}]{deng2009imagenet}
\begin{bchapter}
\bauthor{\bsnm{Deng}, \binits{J.}},
\bauthor{\bsnm{Dong}, \binits{W.}},
\bauthor{\bsnm{Socher}, \binits{R.}},
\bauthor{\bsnm{Li}, \binits{L.-J.}},
\bauthor{\bsnm{Li}, \binits{K.}},
\bauthor{\bsnm{Fei-Fei}, \binits{L.}}:
\bctitle{Imagenet: A large-scale hierarchical image database}.
In: \bbtitle{2009 IEEE Conference on Computer Vision and Pattern Recognition}
(\byear{2010}).
\doiurl{10.1109/cvpr.2009.5206848}
\end{bchapter}
\endbibitem

\bibitem[\protect\citeauthoryear{Serengil and Ozpinar}{2021}]{serengil2021hyperextended}
\begin{bchapter}
\bauthor{\bsnm{Serengil}, \binits{S.I.}},
\bauthor{\bsnm{Ozpinar}, \binits{A.}}:
\bctitle{Hyperextended lightface: A facial attribute analysis framework}.
In: \bbtitle{2021 International Conference on Engineering and Emerging Technologies (ICEET)}
(\byear{2021}).
\doiurl{10.1109/ICEET53442.2021.9659697}
\end{bchapter}
\endbibitem

\bibitem[\protect\citeauthoryear{Grootendorst}{2022}]{grootendorst2022bertopic}
\begin{botherref}
\oauthor{\bsnm{Grootendorst}, \binits{M.}}:
Bertopic: Neural topic modeling with a class-based tf-idf procedure.
https://arxiv.org/abs/2203.05794
(2022)
\end{botherref}
\endbibitem

\bibitem[\protect\citeauthoryear{McInnes et~al.}{2017}]{mcinnes2017hdbscan}
\begin{botherref}
\oauthor{\bsnm{McInnes}, \binits{L.}},
\oauthor{\bsnm{Healy}, \binits{J.}},
\oauthor{\bsnm{Astels}, \binits{S.}}:
hdbscan: Hierarchical density based clustering.
The Journal of Open Source Software
\textbf{2}
(2017)
\doiurl{10.21105/joss.00205}
\end{botherref}
\endbibitem

\bibitem[\protect\citeauthoryear{McInnes et~al.}{2018}]{mcinnes2018umap}
\begin{botherref}
\oauthor{\bsnm{McInnes}, \binits{L.}},
\oauthor{\bsnm{Healy}, \binits{J.}},
\oauthor{\bsnm{Saul}, \binits{N.}},
\oauthor{\bsnm{Großberger}, \binits{L.}}:
Umap: Uniform manifold approximation and projection.
Journal of Open Source Software
\textbf{3}
(2018)
\doiurl{10.21105/joss.00861}
\end{botherref}
\endbibitem

\bibitem[\protect\citeauthoryear{Hamilton et~al.}{2017}]{hamilton2017inductive}
\begin{bchapter}
\bauthor{\bsnm{Hamilton}, \binits{W.L.}},
\bauthor{\bsnm{Ying}, \binits{R.}},
\bauthor{\bsnm{Leskovec}, \binits{J.}}:
\bctitle{Inductive representation learning on large graphs}.
In: \bbtitle{Neural Information Processing Systems},
vol. \bseriesno{2017-December}
(\byear{2017})
\end{bchapter}
\endbibitem

\bibitem[\protect\citeauthoryear{Manning et~al.}{2014}]{manning2014stanford}
\begin{bchapter}
\bauthor{\bsnm{Manning}, \binits{C.D.}},
\bauthor{\bsnm{Surdeanu}, \binits{M.}},
\bauthor{\bsnm{Bauer}, \binits{J.}},
\bauthor{\bsnm{Finkel}, \binits{J.}},
\bauthor{\bsnm{Bethard}, \binits{S.J.}},
\bauthor{\bsnm{McClosky}, \binits{D.}}:
\bctitle{The stanford corenlp natural language processing toolkit}.
In: \bbtitle{Proceedings of 52nd Annual Meeting of the Association for Computational Linguistics: System Demonstrations},
vol. \bseriesno{2014-June}
(\byear{2014}).
\doiurl{10.3115/v1/p14-5010}
\end{bchapter}
\endbibitem

\bibitem[\protect\citeauthoryear{Hutto and Gilbert}{2014}]{hutto2014vader}
\begin{bchapter}
\bauthor{\bsnm{Hutto}, \binits{C.J.}},
\bauthor{\bsnm{Gilbert}, \binits{E.}}:
\bctitle{Vader: A parsimonious rule-based model for sentiment analysis of social media text}.
In: \bbtitle{Proceedings of the International AAAI Conference on Web and Social Media}
(\byear{2014}).
\doiurl{10.1609/icwsm.v8i1.14550}
\end{bchapter}
\endbibitem

\bibitem[\protect\citeauthoryear{Aloufi et~al.}{2017}]{aloufi2017prediction}
\begin{botherref}
\oauthor{\bsnm{Aloufi}, \binits{S.}},
\oauthor{\bsnm{Zhu}, \binits{S.}},
\oauthor{\bsnm{Saddik}, \binits{A.E.}}:
On the prediction of flickr image popularity by analyzing heterogeneous social sensory data.
Sensors (Switzerland)
\textbf{17}
(2017)
\doiurl{10.3390/s17030631}
\end{botherref}
\endbibitem

\bibitem[\protect\citeauthoryear{Jollife and Cadima}{2016}]{jolliffe2016principal}
\begin{botherref}
\oauthor{\bsnm{Jollife}, \binits{I.T.}},
\oauthor{\bsnm{Cadima}, \binits{J.}}:
Principal component analysis: A review and recent developments
(2016).
\doiurl{10.1098/rsta.2015.0202}
\end{botherref}
\endbibitem

\bibitem[\protect\citeauthoryear{Srivastava et~al.}{2014}]{Srivastava2014}
\begin{botherref}
\oauthor{\bsnm{Srivastava}, \binits{N.}},
\oauthor{\bsnm{Hinton}, \binits{G.}},
\oauthor{\bsnm{Krizhevsky}, \binits{A.}},
\oauthor{\bsnm{Sutskever}, \binits{I.}},
\oauthor{\bsnm{Salakhutdinov}, \binits{R.}}:
Dropout: a simple way to prevent neural networks from overfitting. journal of machine learning research.
Journal of Machine Learning Research
\textbf{15}
(2014)
\end{botherref}
\endbibitem

\bibitem[\protect\citeauthoryear{Wu et~al.}{2019}]{wu2019smp}
\begin{bchapter}
\bauthor{\bsnm{Wu}, \binits{B.}},
\bauthor{\bsnm{Liu}, \binits{B.}},
\bauthor{\bsnm{Cheng}, \binits{W.H.}},
\bauthor{\bsnm{Zeng}, \binits{Z.}},
\bauthor{\bsnm{Liu}, \binits{P.}},
\bauthor{\bsnm{Luo}, \binits{J.}}:
\bctitle{Smp challenge: An overview of social media prediction challenge 2019}.
In: \bbtitle{MM 2019 - Proceedings of the 27th ACM International Conference on Multimedia}
(\byear{2019}).
\doiurl{10.1145/3343031.3356084}
\end{bchapter}
\endbibitem

\bibitem[\protect\citeauthoryear{Wu et~al.}{2017}]{wu2017sequential}
\begin{bchapter}
\bauthor{\bsnm{Wu}, \binits{B.}},
\bauthor{\bsnm{Cheng}, \binits{W.H.}},
\bauthor{\bsnm{Zhang}, \binits{Y.}},
\bauthor{\bsnm{Huang}, \binits{Q.}},
\bauthor{\bsnm{Li}, \binits{J.}},
\bauthor{\bsnm{Mei}, \binits{T.}}:
\bctitle{Sequential prediction of social media popularity with deep temporal context networks}.
In: \bbtitle{IJCAI International Joint Conference on Artificial Intelligence},
vol. \bseriesno{0}
(\byear{2017}).
\doiurl{10.24963/ijcai.2017/427}
\end{bchapter}
\endbibitem

\bibitem[\protect\citeauthoryear{Ding et~al.}{2019}]{ding2019social}
\begin{bchapter}
\bauthor{\bsnm{Ding}, \binits{K.}},
\bauthor{\bsnm{Wang}, \binits{R.}},
\bauthor{\bsnm{Wang}, \binits{S.}}:
\bctitle{Social media popularity prediction: A multiple feature fusion approach with deep neural networks}.
In: \bbtitle{Proceedings of the 27th ACM International Conference on Multimedia}
(\byear{2019}).
\doiurl{10.1145/3343031.3356062}
\end{bchapter}
\endbibitem

\bibitem[\protect\citeauthoryear{}{2023}]{hsu2023gradient}
\begin{botherref}
Gradient boost tree network based on extensive feature analysis for popularity prediction of social posts.
In: Proceedings of the 31st ACM International Conference on Multimedia,
pp. 9451--9455
(2023).
\doiurl{10.1145/3581783.3612843}
\end{botherref}
\endbibitem

\bibitem[\protect\citeauthoryear{Mao et~al.}{2023}]{mao2023enhanced}
\begin{bchapter}
\bauthor{\bsnm{Mao}, \binits{S.}},
\bauthor{\bsnm{Xi}, \binits{W.}},
\bauthor{\bsnm{Yu}, \binits{L.}},
\bauthor{\bsnm{L{\"u}}, \binits{G.}},
\bauthor{\bsnm{Xing}, \binits{X.}},
\bauthor{\bsnm{Zhou}, \binits{X.}},
\bauthor{\bsnm{Wan}, \binits{W.}}:
\bctitle{Enhanced catboost with stacking features for social media prediction}.
In: \bbtitle{Proceedings of the 31st ACM International Conference on Multimedia},
pp. \bfpage{9430}--\blpage{9435}
(\byear{2023}).
\doiurl{10.1145/3581783.3612839}
\end{bchapter}
\endbibitem

\bibitem[\protect\citeauthoryear{Spearman}{1904}]{Spearman1904}
\begin{botherref}
\oauthor{\bsnm{Spearman}, \binits{C.}}:
The proof and measurement of association between two things.
The American Journal of Psychology
\textbf{15}
(1904)
\doiurl{10.2307/1412159}
\end{botherref}
\endbibitem

\end{thebibliography}
\end{document}